\documentclass[prb,aps,english,twocolumn,notitlepage,superscriptaddress]{revtex4-2}

\usepackage{graphicx}
\usepackage{dcolumn}
\usepackage{rotating}
\usepackage{lipsum}
\usepackage{xcolor}
\usepackage{bm}
\usepackage[T1]{fontenc}  
\usepackage{multirow,amsmath,amssymb}
 \usepackage{babel} 
 \usepackage{txfonts}
 \usepackage{epsfig,epsf,psfrag} 
\usepackage{graphicx} 
\usepackage{pslatex}
\usepackage{epic,eepic} 
\usepackage{color,pstcol}
\usepackage{pstricks} 
\usepackage{fancyhdr} 
\usepackage{tabularx}
\usepackage{physics}
\usepackage{url}
\usepackage{listings}
\usepackage{color} 
\usepackage{caption}
\usepackage{ulem}
\usepackage{balance}
\usepackage{subcaption}
\usepackage{float}
\usepackage[utf8]{inputenc}
\usepackage{comment}
\usepackage{siunitx}
\usepackage{makecell}

\usepackage{hyperref}
\hypersetup{
 colorlinks=true,
 linkcolor=blue,
 anchorcolor = blue,
 citecolor = blue,
 filecolor = blue,
 urlcolor = blue
}
\usepackage{dsfont}

\graphicspath{{Images/}}

\def \be {\begin{equation}} 
\def \ee {\end{equation}}

\begin{document}

\preprint{APS/123-QED}

\title{Negative Spin $\Delta_T$ noise Induced by Spin-Flip Scattering and Andreev Reflection}

\title{Negative Spin $\Delta_T$ noise Induced by Spin-Flip Scattering and Andreev Reflection}
\author{Sachiraj Mishra}%
\email{sachiraj29mishra@gmail.com}

\author{Colin Benjamin}%
\email{colin.nano@gmail.com}
\affiliation{School of Physical Sciences, National Institute of Science Education and Research, HBNI, Jatni-752050, India}
\affiliation{Homi Bhabha National Institute, Training School Complex, AnushaktiNagar, Mumbai, 400094, India }

\begin{abstract}
We study charge $\Delta_T$ noise, followed by an examination of spin $\Delta_T$ noise, in the normal metal–spin flipper–normal metal–insulator–superconductor (N–sf–N–I–S) junction. Our analysis reveals a key contrast: while charge $\Delta_T$ noise remains strictly positive, spin $\Delta_T$ noise undergoes a sign reversal from positive to negative, driven by the interplay between spin-flip scattering as well as Andreev reflection. In contrast, charge quantum shot noise remain positive and sign-definite, which is valid for spin quantum shot noise also. The emergence of negative spin $\Delta_T$ noise has two major implications. First, it establishes a clear distinction between spin resolved $\Delta_T$ noise and quantum shot noise: the former is dominated by opposite-spin correlations, whereas the latter is led by same-spin correlations. Second, it provides access to scattering mechanisms that are not captured by quantum shot noise alone. Thus, negative spin $\Delta_T$ noise serves as a unique probe of the cooperative effects of Andreev reflection and spin flipping. We further place our results in context by comparing them with earlier reports of negative $\Delta_T$ noise in strongly correlated systems, such as fractional quantum Hall states, and in multiterminal hybrid superconducting junctions. Overall, this work offers new insights into the mechanisms governing sign reversals in $\Delta_T$ noise and highlights their role as distinctive fingerprints of spin-dependent scattering in superconducting hybrid devices.
\end{abstract}

\maketitle

\section{Introduction} 

The study of $\Delta_T$ noise has recently attracted considerable attention from both theoretical~\cite{PhysRevLett.127.136801, PhysRevLett.125.086801, popoff2022scattering} and experimental~\cite{lumbroso2018electronic, shein2022electronic, sivre2019electronic, PhysRevLett.125.106801, melcer2022absent} researchers, due to its promise as an effective diagnostic probe in nonequilibrium quantum transport. This form of noise persists even when there is no net current, provided there is a finite temperature gradient across the system~\cite{lumbroso2018electronic, PhysRevLett.127.136801, PhysRevB.107.075409}. Experimentally, $\Delta_T$ noise has been observed in a variety of platforms, including atomic-scale molecular junctions~\cite{lumbroso2018electronic}, mesoscopic quantum circuits~\cite{shein2022electronic}, and metallic tunnel junctions~\cite{sivre2019electronic}. Conceptually, $\Delta_T$ noise represents a quantum contribution to shot noise driven solely by thermal gradients, offering a unique window into the nonequilibrium dynamics of charge and spin carriers in mesoscopic systems.

Recent advances have extended the study of $\Delta_T$ noise to superconducting hybrid junctions, as demonstrated by several theoretical works~\cite{mishra2025delta_t, PhysRevResearch.7.023321, k95y-7zrb}. In our recent study \cite{k95y-7zrb}, we showed that charge $\Delta_T$ noise together with spin $\Delta_T$ noise can act as highly sensitive probes for identifying distinct categories of bound states in superconducting systems such as Yu–Shiba–Rusinov (YSR) bound states and Majorana bound states (MBS), thereby providing a diagnostic method that goes beyond conventional tunnelling spectroscopy. While charge (spin) $\Delta_T$ noise exhibits pronounced maxima or minima in regimes where Yu-Shiba-Rusinov states are present, no such signature is observed for MBS \cite{k95y-7zrb}.

\begingroup

\( \Delta_T \) noise can be used to distinguish different edge-mode transport mechanisms such as chiral, helical, and trivial edge modes \cite{mishra2025deltatnoiserobustdiagnostic}, to analyze the role of electron--hole symmetry breaking and Andreev reflection in superconducting hybrid systems \cite{mishra2026deltatnoiseelectronholeasymmetry}, and to probe pairing symmetries in unconventional superconductors such as iron-pnictides \cite{dora2026deltatnoisequantumshot}. Historically, shot noise has been used to study many exotic phenomena in mesoscopic physics. While conductance measurements already encode valuable information about superconducting states, noise properties provide a more sensitive probe of underlying transport mechanisms \cite{noise}. Historically, quantum shot noise has played a central role in mesoscopic superconductivity, including the detection of effective charge \( 2e \) in normal metal--superconductor junctions \cite{jehl2000detection, kozhevnikov2000shot, PhysRevLett.84.3398}. More generally, shot noise has been widely used to probe fractional charge and anyonic statistics in fractional quantum Hall systems \cite{PhysRevLett.79.2526, de1998direct, reznikov1999observation, PhysRevLett.91.216804}, to probe topological transport in quantum Hall setups \cite{mani2017probing, PhysRevB.108.115301}, to detect Majorana bound states \cite{PhysRevB.91.081405, PhysRevB.107.155416, k95y-7zrb}, and to study entanglement in Cooper-pair splitting devices \cite{hofstetter2009cooper}.

\endgroup

A recent study~\cite{PhysRevLett.125.086801} on strongly correlated systems, specifically a two-terminal fractional quantum Hall setup with a quantum point contact, reported the emergence of negative charge $\Delta_T$ noise. This counterintuitive behavior was attributed to the tunnelling of fractionalized Laughlin quasiparticles between terminals, an effect absent when only electron tunnelling is considered. Along similar lines, Ref.\cite{PhysRevB.105.195423} linked the sign change in $\Delta_T$ noise to the exchange statistics of tunnelling quasiparticles, while Ref.\cite{PhysRevResearch.7.023321} demonstrated a sign reversal in cross-correlated $\Delta_T$ noise in a multiterminal superconducting–integer quantum Hall platform. In sharp contrast to these works, which rely on exotic strongly correlated quasiparticles and edge-mode transport, our study reveals that spin $\Delta_T$ noise can undergo sign reversal purely from the synergy of spin-flip scattering as well as Andreev reflection in the ballistic regime of N–sf–N–I–S junction. This represents a fundamentally new finding: the first observation of negative spin $\Delta_T$ noise without fractionalization or topological edge modes, offering a simpler and more experimentally accessible pathway to spin-resolved $\Delta_T$ noise.
  
\begingroup
    
   In the present study, we demonstrate that the interplay between spin-flip scattering and Andreev reflection governs both the magnitude and the sign of spin \( \Delta_T \) noise, leading to a temperature-driven sign reversal that has not been identified previously. Unlike earlier studies, which focused on charge \( \Delta_T \) noise in strongly correlated systems and complex hybrid superconducting junctions, our results show that such sign changes can arise in a simple superconducting hybrid junction and originate from a crossover between Andreev-reflection-dominated and quasiparticle-dominated transport. We further establish a clear conceptual distinction between spin \( \Delta_T \) noise and spin quantum shot noise, demonstrate that the observed sign reversal represents a genuine qualitative change in correlation, and identify the relevant temperature and parameter regimes where this transition occurs. These effects are shown to be robust over a wide range of parameters and address an important gap in the literature, namely the lack of a unified understanding of how spin-flip scattering and Andreev reflection jointly influence spin-resolved \( \Delta_T \) noise.
\endgroup

The structure of the paper is organized as: in Sec.~\ref{theory}, we first present a brief overview of our chosen N–sf–N–I–S setup. We then explain the phenomena of spin-flip scattering by introducing a spin flipper at the metal/metal interface and an insulator at the metal–superconductor interface. Next, we discuss our results of charge $\Delta_T$ noise, followed by spin $\Delta_T$ noise, together with the corresponding charge as well as spin quantum shot noise in the considered setup. In Section~\ref{analysis}, we present a comprehensive analysis of our findings, highlighting charge $\Delta_T$ noise followed by spin $\Delta_T$ noise, in comparison with the corresponding quantum shot noise. We also systematically compare our findings with previous studies \cite{PhysRevLett.125.086801, PhysRevB.105.195423, PhysRevResearch.7.023321} that reported negative charge $\Delta_T$ noise. In Sec. \ref{real}, we disuss the possible experimental realization of our work. We conclude in Section~\ref{expt} with a discussion on summary of our findings. The MATHEMATICA code used for calculating $\Delta_T$ noise and quantum shot noise is available in Ref. \cite{github}. The MATHEMATICA code used for calculating $\Delta_T$ noise and quantum shot noise is available in Ref. \cite{github}.

\section{Theory}
\label{theory}

\begin{figure}[H]
\centering
\includegraphics[width=1.00\linewidth]{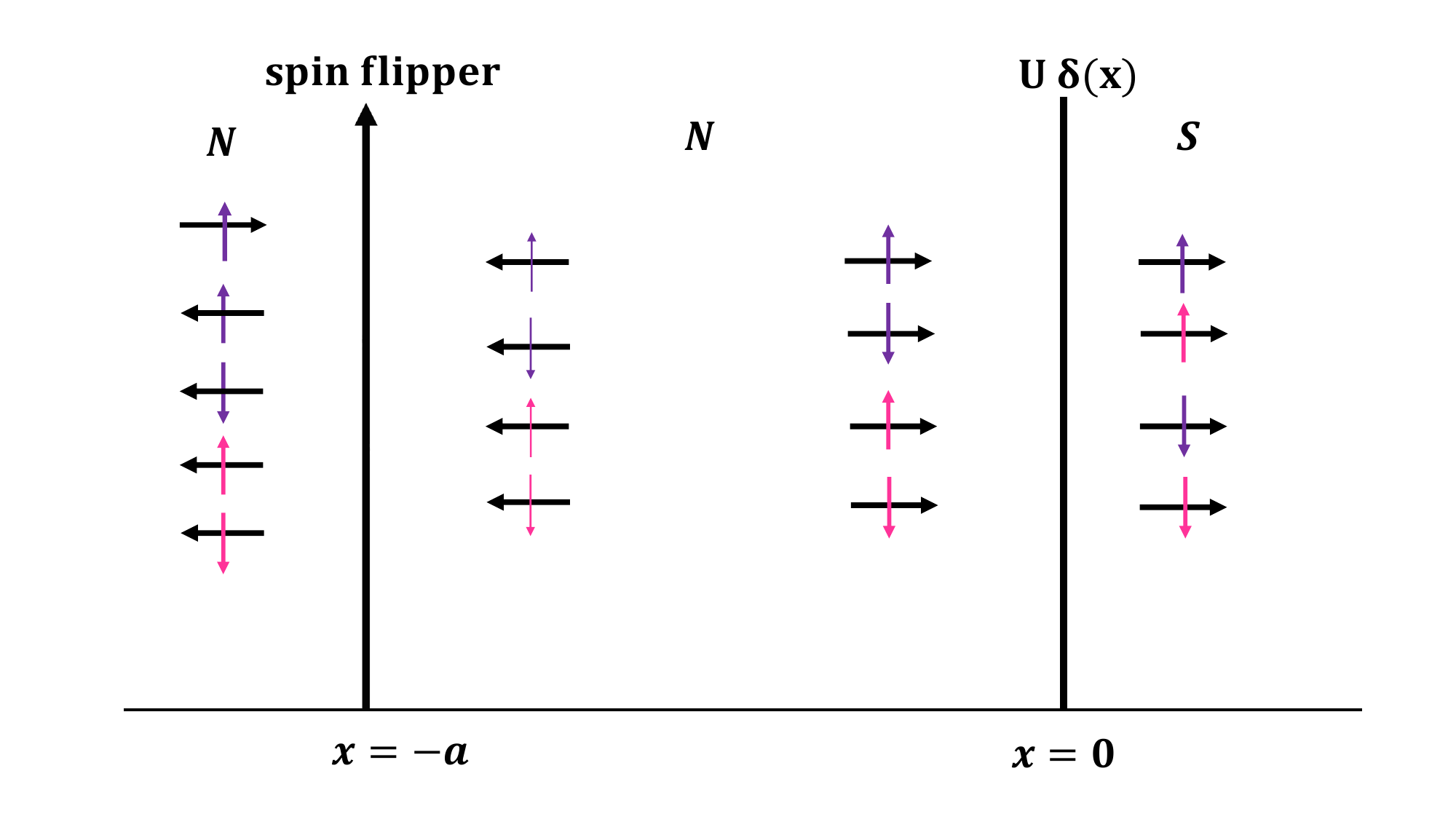}
\caption{Schematic illustration of a one-dimensional N–sf–N–I–S junction, with a localized magnetic impurity situated at $x=-a$ serves as a spin-flipper, and the insulating barrier at $x=0$ models the N–S interface. Purple $\uparrow$ ($\downarrow$) arrows represent spin-up (spin-down) electrons, whereas pink $\uparrow$ ($\downarrow$) arrows correspond to spin-up (spin-down) holes.}
\label{fig:1}
\end{figure}

In Fig.~\ref{fig:1}, a one-dimensional N-sf-N-I-S junction consisting of a spin magnetic impurity at the interface $x=-a$ is shown. When an electron with spin either in up or down direction is incident from left normal metal $N_1$ on the interfacial spin-flipper ($x=-a$), there can be mutual spin-flip, and the electron may be normally reflected or Andreev reflceted to normal metal $N_1$ or transmitted to superconductor both as electron-like quasiparticle and hole-like quasiparticle. The BdG Hamiltonian describing the various scattering events occurring within the N–sf–N–I–S setup is written as

\begin{equation}
    \mathcal{H}=\left(
	\begin{array}{cc}
	(H_0 (k) + 	H_{sf}) \hat{I} & \hat{\Delta}(\textbf{k}) \Theta(x-a)  \\
		-\hat{\Delta}^*(-\textbf{k}) \Theta(x-a)  & -(H_0^* (-k) + 	H_{sf}^*) \hat{I}
	\end{array}
	\right).
	\label{eq1}
\end{equation}

We consider the single-particle Hamiltonian  
$H_0(k) = \hbar^2 k^2/2m^* + U \delta(x) - E_F$,  
with $k$ representing the wave vector, $E_F$ the Fermi energy, and $m^*$ the quasiparticle's effective mass. The term $U \delta(x)$ models the insulating barrier through a delta-function potential. The effect of a localized magnetic impurity at the interface is incorporated via the spin-flip Hamiltonian
$H_{sf} = -J_0 \delta(x),\vec{s}\cdot\vec{\Sigma}$,
with $J_0$ quantifying the interaction strength linking the quasiparticle spin $\vec{s}$ with the impurity spin $\vec{\Sigma}$. Regarding superconductivity, the pairing potential is $\hat{\Delta}(\mathbf{k})$ (Refs.~\cite{PhysRevB.106.125402, RevModPhys.88.035005, PhysRevB.78.195125}). In the case of conventional $s$-wave singlet pairing, the gap takes the form $\hat{\Delta}(\mathbf{k}) = i\Delta_0 \chi_y$, where $\chi_i$ $(i \in \{x,y,z\})$ denote the Pauli matrices. The exchange interaction present in $H_{sf}$ is: 
\begin{eqnarray}
    \vec{s} \cdot \vec{\Sigma} = s_z \cdot \Sigma_z + \frac{1}{2} (s^- \Sigma^+ + s^+ \Sigma^-).
    \label{eq2}
\end{eqnarray}

The electron spin operators are denoted by $s_x, s_y, s_z$, whereas the spin-flipper (impurity) spin components are represented by $\Sigma_x, \Sigma_y, \Sigma_z$. The corresponding ladder operators are defined in the usual way as $s^{\pm} = s_x \pm i s_y$ for the electron spin and $\Sigma^{\pm} = \Sigma_x \pm i \Sigma_y$ for the spin-flipper spin.

\subsection{Spin-flip scattering}
\label{spinflip}

\begin{widetext}

The wave functions in all the three regions considered for spin-up electron incident from $N_1$ can be written as (see, Fig.~\ref{fig:1}),
\begin{eqnarray}
\phi_{N_1}(x) &=& \left[ e^{i q_e x } + b_{\uparrow\uparrow} e^{-i q_e x } \right] \Psi^{s}_{m} \chi_1 
+ b_{\downarrow\uparrow} e^{-i q_e x} \Psi^{s}_{m+1} \chi_2 
+ a_{\uparrow\uparrow} e^{i q_h x} \Psi^{s}_{m+1} \chi_3 
+ a_{\downarrow\uparrow} e^{i q_h x} \Psi^{s}_{m} \chi_4, 
\quad x<-a, \nonumber \\
\phi_{N_2}(x) &=& t_{\uparrow\uparrow} e^{i q_e x } \Psi^{s}_{m} \chi_1 
+ t_{\downarrow\uparrow} e^{i q_e x} \Psi^{s}_{m+1} \chi_2 
+ f_{\uparrow\uparrow} e^{-i q_e (x-a) } \Psi^{s}_{m} \chi_1  
+ f_{\downarrow\uparrow} e^{-i q_e (x-a)} \Psi^{s}_{m+1} \chi_2 \nonumber \\ 
&& + g_{\uparrow\uparrow} e^{i q_h (x-a)} \Psi^{s}_{m+1} \chi_3  
+ g_{\downarrow\uparrow} e^{i q_h (x-a)} \Psi^{s}_{m} \chi_4  
+ h_{\uparrow\uparrow} e^{-i q_h x} \Psi^{s}_{m+1} \chi_3  
+ h_{\downarrow\uparrow} e^{-i q_h x} \Psi^{s}_{m} \chi_4,  
\quad -a<x<0, \nonumber \\
\phi_{S}(x) &=& c_{\uparrow\uparrow} e^{i k_e x } \Psi^{s}_{m} \chi^S_1 
+ c_{\downarrow\uparrow} e^{i k_e x} \Psi^{s}_{m+1} \chi^S_2 
+ d_{\uparrow\uparrow} e^{-i k_h x} \Psi^{s}_{m+1} \chi^S_3 
+ d_{\downarrow\uparrow} e^{-i k_h x} \Psi^{s}_{m} \chi^S_4,  
\quad x>0.
\label{eq1}
\end{eqnarray}
\end{widetext}

where $\Psi^s_m$ is the eigenfunction of $s_z$ such that $s_z \Psi^s_m = m \Psi^s_m$, with $m$ being spin integer number and $k_e$ being wave vector of electron, i.e., $k_e = \sqrt{\frac{2 m^*}{\hbar^2} ( E_F + \sqrt{E^2 - \Delta_0^2} )} = k_F \sqrt{( 1 + \frac{\sqrt{E^2 - \Delta_0^2}}{E_F} )} \simeq k_F$ with energy of incident electron $E>0$, and $k_F=\sqrt{\frac{2 m^* E_F}{\hbar^2}}$ is Fermi wave vector with $E_F$ being the Fermi energy. $\chi_i$ is the spinor in normal metal and $\chi_i^S$ in superconductor. However, the superconducting gap $\Delta$ changes with temperature, which is equal to $\Delta (T) = \Delta_0 \sqrt{1 - \frac{T}{T_C}}$, where $\Delta_0 = 1.76*k_B T_C$, with $T_C = 9K$ for Nb \cite{PhysRev.95.1435, sun2023smooth}. More details of the calculation is given in Ref. \cite{k95y-7zrb}. Reflection amplitudes for an incident spin-up electron to be reflected as spin-up electron is denoted as $b_{\uparrow \uparrow}$ and to be reflected as spin-down electron is denoted as $b_{\downarrow \uparrow}$. Similarly, $a_{\uparrow \uparrow}$ being the Andreev reflection amplitude with same spin, whereas $a_{\downarrow \uparrow}$ being the Andreev reflection amplitudes with different spins. In a similar manner, $c_{\uparrow \uparrow}$ denotes the transmission coefficient corresponding to an incoming spin-up electron that propagates into the superconductor as a spin-up electron-like quasiparticle. On the other hand, $c_{\downarrow \uparrow}$ represents the transmission amplitude for spin-up electron incident from the normal side to emerge as a spin-down electron-like quasiparticle. Likewise, $d_{\uparrow \uparrow}$ and $d_{\downarrow \uparrow}$ describe the conversion of the incoming electron into hole-like quasiparticles in the superconducting region, with $d_{\uparrow \uparrow}$ corresponding to the same-spin process and $d_{\downarrow \uparrow}$ corresponding to the opposite-spin process.

{The electron spin is represented by the operator $\vec{s}$, while the spin-flipper's spin is denoted by $\vec{\Sigma}$. When these operators act on the spinor of an up-spin electron along with the spin-flipper state, they yield the following relations~\cite{de1984spin, pal2019spin, pal2018yu}:
}
{
\begin{eqnarray}
	\Vec{s}. \Vec{\Sigma} \left( \begin{array}{c}
		1\\
		0\\
		0\\
		0
	\end{array} \right) \Psi^s_m = \frac{m}{2} \left( \begin{array}{c}
		1\\
		0\\
		0\\
		0
	\end{array} \right) \Psi^s_{m} + \frac{\tau}{2} \left( \begin{array}{c}
		0\\
		1\\
		0\\
		0
	\end{array} \right) \Psi^s_{m+1}.
	\label{eq11}
\end{eqnarray}}

In this framework, $\tau = \sqrt{(\Sigma - m)(\Sigma + m + 1)}$ and $\tau_1 = \sqrt{(\Sigma + m)(\Sigma - m + 1)}$ correspond to the probabilities of spin flip for incident electrons having spin state up and down, respectively, from the left normal lead.
Here, $\Sigma$ stands for the spin-flipper's spin, while $m$ takes integer values in the range $-\Sigma, -\Sigma + 1, \dots, \Sigma - 1, \Sigma$.
Similarly, one can also see the similar equations involving the operation of $\Vec{s}. \Vec{\Sigma}$ on the spinor of spin-down electron, spin-up hole and spin-down hole, see Ref. \cite{k95y-7zrb}.

The boundary matching conditions at the interfaces of the considered setup can be expressed as follows. At the position of the magnetic impurity, $x=-a$, we have
\begin{eqnarray}
\phi_{N_1}(-a) &=& \phi_{N_2}(-a), \nonumber \\
\frac{d \phi_{N_2}( -a)}{dx} - \frac{d \phi_{N_1}(-a)}{dx} &=& - \frac{2 m^* J_0}{\hbar^2} \, (\vec{s} \cdot \vec{\Sigma}) \, \phi_{N_1}(-a),
\end{eqnarray}
where the exchange interaction occurs due to the spin-flipper. At the normal metal–superconductor interface, $x=0$, the boundary conditions are
\begin{eqnarray}
\phi_{N_2}(0) &=& \phi_{S}(0), \nonumber \\
\frac{d \phi_{S}(0)}{dx} - \frac{d \phi_{N_2}(0)}{dx} &=& \frac{2 m^* U}{\hbar^2} \, \phi_{N_2}(0),
\end{eqnarray}
where $U$ denotes the strength of the insulating barrier. Substituting the wavefunctions defined in Eq.~(\ref{eq1}) into these boundary conditions allows us to solve for the scattering amplitudes corresponding to an incident electron with either spin orientation. In our calculations of quantum shot noise and $\Delta_T$ noise, we use the dimensionless barrier strength $Z = \frac{m^* U}{\hbar^2 k_F}$ and the dimensionless exchange interaction $J = \frac{m^* J_0}{\hbar^2 k_F}$.

\begin{figure}[h!]
\begin{center}
\boxed{
\begin{aligned}
1. \big\uparrow_{e^-} & \otimes \big\Uparrow_S \xrightarrow[]{\text{ $S=m$ }}
\frac{m}{2} \left( \big\uparrow_{e^-} \otimes \big\Uparrow_S \right) \nonumber \\
2. \big\uparrow_{e^-} & \otimes \big\Downarrow_S \xrightarrow[]{\text{ $S \neq m$ }}
\frac{m}{2} \left( \big\uparrow_{e^-} \otimes \big\Downarrow_S \right) + \frac{\tau}{2} \left( \big\downarrow_{e^-} \otimes \big\Uparrow_S \right)  \nonumber \\
3. \big\downarrow_{e^-} & \otimes \big\Uparrow_S \xrightarrow[]{\text{ $S\neq-m$ }}
\frac{-m}{2} \left( \big\downarrow_{e^-} \otimes \big\Uparrow_S \right) + \frac{\tau_1}{2} \left( \big\uparrow_{e^-} \otimes \big\Downarrow_S \right)  \nonumber \\
4. \big\downarrow_{e^-} & \otimes \big\Downarrow_S \xrightarrow[]{\text{ $S = -m$ }}
\frac{-m}{2} \left( \big\downarrow_{e^-} \otimes \big\Downarrow_S \right)
\end{aligned} }
\end{center}
\caption{Illustration of different spin configurations when spin $\uparrow$ or $\downarrow$ electron encounters a spin-flipper with $S= \pm m$ or $S \neq \pm m$. $\tau$ and $\tau_1$ are spin-flip probabilities for spin-up and spin-down incident electrons. }
\label{fig:avgm}
\end{figure}
One can also consider four different possible configurations as shown in Fig. \ref{fig:avgm}. When the electron's elastic scattering time is much greater than the relaxation time of the spin-flipper $\tau_e \gg \tau_{sf}$, the spin-flipper will flip back before it interacts with the upcoming incident electron, see Fig.~\ref{fig:avgm}. Therefore, spin magnetic moment $m$ for a specific spin-flipper's spin $S$ is not fixed, and to calculate any transport quantity, the average over all possible $m$ values is taken. We consider the cases where a spin-up incident electron interacts in either spin-configuration 1 ($S=m$) or spin-configuration 2 ($S \neq m$). Likewise, for spin-down incident electron, the spin-flipper interacts via spin-configuration 3 ($S \neq -m$) or spin-configuration 4 ($S=-m$), as shown in Fig. \ref{fig:avgm}.

In the following subsection, we compute the charge (spin) currents, along with the quantum charge (spin) quantum shot noise followed by $\Delta_T$ noise, arising from spin-flip scattering in our setup depicted in Fig.~\ref{fig:1}.

\subsection{Charge and spin current}
\label{current}

For the setup shown in Fig.~\ref{fig:1}, the average spin-polarized current can be expressed as~\cite{PhysRevB.53.16390,noise}  
\begin{eqnarray}
\langle I^{s}_i \rangle = \frac{e}{h} \sum_{\substack{j,k \in \{1,2\} \\ \alpha,\beta,\gamma \in \{e,h\} \\ \sigma, \sigma' \in \{\uparrow, \downarrow\}}} \text{sgn}(\alpha) \int_{-\infty}^{\infty} dE \, \mathcal{M}^{\sigma \sigma'}_{jk;\beta\gamma}(i \alpha, s) \, \langle b^{\sigma \dagger}_{j \beta} b^{\sigma'}_{k \gamma} \rangle,
\label{spi_pol}
\end{eqnarray}
where $\text{sgn}(\alpha) = +1$ ($-1$) for electrons (holes). The matrix $\mathcal{M}$ is defined as
\begin{equation}
\mathcal{M}^{\sigma \sigma'}_{jk;\beta\gamma}(i \alpha, s) = \delta_{ij} \delta_{ik} \delta_{\alpha \beta} \delta_{\alpha \gamma} \delta_{s \sigma} \delta_{s \sigma'} - S^{\alpha \beta, s \sigma \dagger}_{ij} S^{\alpha \gamma, s \sigma'}_{ik},
\end{equation}
with $i,j,k \in \{1,2\}$ denoting the contacts (left normal lead and superconducting lead), $\alpha, \beta, \gamma$ labeling electron and holes. Here, $b^{\sigma \dagger}_{j \beta}$ ($b^{\sigma'}_{k \gamma}$) is the creation (annihilation) operator for a particle of kind $\beta$ ($\gamma$) situated at contact $j$ ($k$) carrying spin $\sigma$ ($\sigma'$). $S_{ij}^{\alpha \beta, s \sigma' }$ is the scattering amplitude for a particle of type $\beta$ with spin $\sigma'$ to scatter from terminal $j$ into terminal $i$ as a particle type $\alpha$ with spin $s$.

{In Eq. (\ref{spi_pol}), $\langle I_i^{s} \rangle$ denotes the {spin-resolved or spin-polarized charge current}
associated with spin $s \in \{\uparrow,\downarrow\}$ flowing into terminal $i$.
By ``spin-polarized current,'' we do {not} mean a net spin current (i.e., the difference
between spin-up and spin-down currents), but rather the charge current carried by particles
with fixed spin orientation $s$.}

{Although Eq.~(\ref{spi_pol}) above contains a summation over the spin indices $\sigma$ and $\sigma'$, these
indices are dummy indices, analogous to the terminal indices $j,k$ and the particle-types $\alpha,\beta,\gamma$. Their summation over these indices accounts for all possible spin-dependent
scattering processes contributing to the spin-polarized current with spin $s$. Spin selectivity
enters through the element: $\mathcal{M}^{\sigma \sigma'}_{jk;\beta\gamma}(i\alpha,s)$ in Eq. (\ref{spi_pol}), 
which is explicitly defined for a fixed outgoing spin $s$. As a result, the current
$\langle I_i^{s} \rangle$ corresponds to the charge current resolved in the spin-$s$ sector or a spin-polarized current with spin $s$.}

{The summations over $\sigma$ or $\sigma'$ do not imply spin averaging of the
measured current; instead, it reflects the internal summation over scattering channels,
while the label $s$ specifies the spin-resolved or spin-polarized current component.}

The average value is given by $\langle b^{\sigma \dagger}_{j \beta} b^{\sigma'}_{k \gamma} \rangle = \delta_{jk} \delta_{\beta \gamma} \delta_{\sigma \sigma'} f_{j \beta}(E)$, where $f_{j \beta}(E)$ is the Fermi distribution in contact $j$ for particle $\beta$, independent of spin. Explicitly, $f_{j \beta}(E) = \left[ 1 + \exp\left(\frac{E + \text{sgn}(\beta) e V_j}{k_B T_j}\right) \right]^{-1}$, where $\text{sgn}(\beta) = \pm 1$ for electrons (holes), $k_B$ being Boltzmann constant with $T_j$ denoting the temperature, whereas $V_j$ is the voltage bias applied at lead $j$.  

{The average spin up polarized charge current for spin
$s=\uparrow$, i.e., for a spin-up incident electron, is given by}
{
\begin{equation}
\langle I^{\uparrow}_1 \rangle
= \frac{e}{h} \int_{-\infty}^{\infty} dE \,
\big( 1 + \mathcal{R}^A_{\uparrow\uparrow}
      + \mathcal{R}^A_{\downarrow\uparrow}
      - \mathcal{R}_{\uparrow\uparrow}
      - \mathcal{R}_{\downarrow\uparrow} \big)
\big( f_{1e}(E) - f_{2e}(E) \big),
\label{eq:Iup}
\end{equation}}
{where $\mathcal{R}^A_{\uparrow\uparrow}=|a_{\uparrow\uparrow}|^2$ is the Andreev reflection
probability without spin flip, $\mathcal{R}^A_{\downarrow\uparrow}=|a_{\downarrow\uparrow}|^2$
is the Andreev reflection probability with spin flip,
$\mathcal{R}_{\uparrow\uparrow}=|b_{\uparrow\uparrow}|^2$ is the normal reflection
probability without spin flip, and
$\mathcal{R}_{\downarrow\uparrow}=|b_{\downarrow\uparrow}|^2$ is the normal reflection
probability with spin flip.}

{Similarly, the average spin down polarized charge current for spin $s=\downarrow$, corresponding
to a spin-down incident electron in the normal metal, is}
{
\begin{equation}
\langle I^{\downarrow}_1 \rangle
= \frac{e}{h} \int_{-\infty}^{\infty} dE \,
\big( 1 + \mathcal{R}^A_{\downarrow\downarrow}
      + \mathcal{R}^A_{\uparrow\downarrow}
      - \mathcal{R}_{\downarrow\downarrow}
      - \mathcal{R}_{\uparrow\downarrow} \big)
\big( f_{1e}(E) - f_{2e}(E) \big),
\label{eq:Idown}
\end{equation}}
{where $\mathcal{R}^A_{\downarrow\downarrow}=|a_{\downarrow\downarrow}|^2$,
$\mathcal{R}^A_{\uparrow\downarrow}=|a_{\uparrow\downarrow}|^2$,
$\mathcal{R}_{\downarrow\downarrow}=|b_{\downarrow\downarrow}|^2$, and
$\mathcal{R}_{\uparrow\downarrow}=|r^{Na}_{\uparrow\downarrow}|^2$ are defined analogously as before.}

{In our setup, due to spin symmetry, the scattering probabilities for spin-up and spin-down
incident electrons are identical, i.e.,
$
\mathcal{R}^A_{\downarrow\downarrow} = \mathcal{R}^A_{\uparrow\uparrow}, \quad
\mathcal{R}^A_{\uparrow\downarrow} = \mathcal{R}^A_{\downarrow\uparrow}, \quad
\mathcal{R}_{\downarrow\downarrow} = \mathcal{R}_{\uparrow\uparrow}, \quad
\mathcal{R}_{\uparrow\downarrow} = \mathcal{R}_{\downarrow\uparrow}.
$. As a consequence, the spin-resolved or spin-polarized charge currents satisfy
$\langle I^{\uparrow}_1 \rangle = \langle I^{\downarrow}_1 \rangle$.}

{Thus, the total average charge current,
defined as the sum of the two spin-polarized contributions,
$\langle I^{ch}_1 \rangle = \langle I^{\uparrow}_1 \rangle + \langle I^{\downarrow}_1 \rangle$, acquires a factor of $2$ and} the mean charge current for $N_1$ as shown in in Fig. \ref{fig:1} can then be written as~\cite{de1984spin, pal2019spin, pal2018yu}  
\begin{equation}
\langle I^{ch}_1 \rangle = \frac{2e}{h} \int_{-\infty}^{\infty} dE \, F^{ch}(E) \, \big( f_{1e}(E) - f_{2e}(E) \big),
\label{eq1000}
\end{equation}
with the function $F^{ch}(E) = 1 + \mathcal{R}^A_{\uparrow \uparrow} + \mathcal{R}^A_{\downarrow \uparrow} - \mathcal{R}_{\uparrow \uparrow} - \mathcal{R}_{\downarrow \uparrow}$. Here, $\mathcal{R}^A_{\uparrow \uparrow} = |a_{\uparrow \uparrow}|^2$ is the Andreev reflection probability with same spin, $\mathcal{R}^A_{\downarrow \uparrow} = |a_{\downarrow \uparrow}|^2$ is the Andreev reflection probability with opposite spin, $\mathcal{R}_{\uparrow \uparrow} = |b_{\uparrow \uparrow}|^2$ is the normal reflection with same spin and $\mathcal{R}_{\downarrow \uparrow} = |r_{\downarrow \uparrow}^{Na}|^2$ is the normal reflection with opposite spin.
 Similarly, mean spin current is given as

{
\begin{equation}
\begin{split}
\langle I^{sp}_1 \rangle
&= \frac{2e}{h} \int^{\infty}_{-\infty} dE \,\,\, F^{sp}(E) (f_{1e}(E)-f_{2e}(E)) ,
\label{eq23}
\end{split}
\end{equation}}

{with $F^{sp} (E) = 1 + \mathcal{R}^A_{\uparrow \uparrow} - \mathcal{R}^A_{\downarrow \uparrow} - \mathcal{R}_{\uparrow \uparrow} + \mathcal{R}_{\downarrow \uparrow}$. Since our setup preserves electron–hole symmetry, the currents (charge as well as spin) vanish when applied voltage bias is zero, even with a finite temperature gradient. This symmetry ensures that the Seebeck coefficient of the hybrid junction is identically zero, implying that the average current is solely governed by the applied voltage. Consequently, the average current becomes zero only when the voltage bias is zero.}

\subsection{Quantum noise}
\label{qnoise}

Quantum noise autocorrelation implies the correlation of current fluctuations within $N_1$ as shown in Fig. \ref{fig:1}, and the formula for spin-polarised charge quantum noise correlation at zero frequency is
\begin{equation}
 Q^{ch}_{11} = Q^{ \uparrow \uparrow}_{11} + Q^{ \uparrow \downarrow}_{11} + Q^{ \downarrow \uparrow}_{11} + Q^{ \downarrow \downarrow}_{11}.
 \label{eq24}
\end{equation}
Similarly, spin quantum noise is
\begin{equation}
    Q^{sp}_{11} = Q^{ \uparrow \uparrow}_{11} - Q^{ \uparrow \downarrow}_{11} - Q^{ \downarrow \uparrow}_{11} + Q^{ \downarrow \downarrow}_{11},
    \label{eq25}
\end{equation},

\begin{widetext}

The spin polarized quantum noise valid at zero frequency in $N_1$ is  
\begin{equation}
\mathcal{Q}^{s s'}_{11} = \sum_{\alpha, \beta \in \{e,h\}} \mathcal{Q}^{s s', \alpha \beta}_{11} = \frac{e^2}{h} \int_{-\infty}^{\infty} dE \sum_{\sigma, \sigma' \in \{\uparrow, \downarrow\}} \sum_{\substack{j,k \in \{1,2\} \\ \alpha, \beta, \gamma, \delta \in \{e,h\}}} \text{sgn}(\alpha) \text{sgn}(\beta) \, \mathcal{M}^{\sigma \sigma'}_{j \gamma; k \delta}(1 \alpha, s) \, \mathcal{M}^{\sigma' \sigma}_{k \delta; j \gamma}(1 \beta, s') \, f_{j \gamma}(E) \big[ 1 - f_{k \delta}(E) \big],
\end{equation}
where $\text{sgn}(\alpha) = +1$ ($-1$) for electrons (holes). The matrix $\mathcal{M}$ is defined as
\begin{equation}
\mathcal{M}^{\sigma \sigma'}_{j \gamma; k \delta}(p \alpha, s) = \delta_{pj} \delta_{pk} \delta_{\alpha \gamma} \delta_{\alpha \delta} \delta_{s \sigma } \delta_{s \sigma'} - S^{\alpha \gamma, s \sigma  \dagger}_{pj} S^{\alpha \delta, s \sigma'}_{pk}.
\end{equation}
The quantity $S^{\alpha \gamma, s \sigma}_{pk}$ characterizes the amplitude representing a particle initially of type $\gamma $ with spin index $\sigma$ at terminal $k$ to emerge at terminal $p$ as particle of kind $\alpha $ with spin $s$, where $\gamma, \alpha \in {e,h}$ and $\sigma, s \in {\uparrow, \downarrow}$.  

 \end{widetext}

The complete expressions for $Q_{11}^{\mathrm{ch}}$ and $Q_{11}^{\mathrm{sp}}$ are provided in Ref.~\cite{k95y-7zrb}, where the Fermi wave vector parameter $k_F a$ was fixed at $0.85\pi$. In contrast, in this work, {we consider Nb as the superconducting material, whose lattice constant (\( a_0 \)) is approximately 3.30040 Angstrom, therefore, the corresponding Fermi wave vector is $k_F = \frac{2\pi}{a_0} \simeq 1.2 \times 10^{10}~\mathrm{m^{-1}}.$ Here, we take \( k_F \approx 10^{10}~\mathrm{m^{-1}} \). The length of the normal metallic region is chosen as \( a = 10~\mathrm{nm} \), leading to \( k_F a = 100 \).}
Therefore, the shot noise-like part of the total quantum noise $Q_{11}^{sh; s s'}$ for $s, s' \in \{\uparrow, \downarrow\}$, which are given as,
\begin{widetext}

\begin{equation}
\begin{split}
Q_{11}^{sh; \uparrow \uparrow} &= Q_{11}^{sh; \downarrow \downarrow} = \frac{2e^2}{h} \int_{-\infty}^{\infty} dE \, (f_{1e} - f_{1h})^2 \bigg[2 \mathcal{R}^A_{\downarrow \uparrow} \mathcal{R}_{\downarrow \uparrow} + 2 \mathcal{R}^A_{\uparrow \uparrow} \mathcal{R}_{\uparrow \uparrow} + \mathcal{R}^A_{\uparrow \uparrow} \mathcal{R}_{\downarrow \uparrow} + \mathcal{R}^A_{\downarrow \uparrow} \mathcal{R}_{\uparrow \uparrow} - 2 Re (a_{\downarrow \uparrow } b_{\uparrow \uparrow} a^*_{\uparrow \uparrow  } b^*_{\downarrow \uparrow  })
\bigg] \\
&\quad + \frac{2e^2}{h} \int_{-\infty}^{\infty} dE \, (f_{1e} - f_{2e})^2 \bigg[(\mathcal{R}^A + \mathcal{R})(\mathcal{C}^S + \mathcal{D}^S)
\bigg], \\[1.2em]
Q_{11}^{sh; \uparrow \downarrow} &= Q_{11}^{sh; \downarrow \uparrow} =\frac{2e^2}{h} \int_{-\infty}^{\infty} dE \, (f_{1e} - f_{1h})^2 \bigg[  4 Re (a_{\uparrow \uparrow} b_{\uparrow \uparrow} a^*_{\downarrow \uparrow } b^*_{\uparrow \uparrow }) - \mathcal{R}^A_{\uparrow \uparrow   } \mathcal{R}_{\downarrow \uparrow} - \mathcal{R}^A_{\downarrow \uparrow} \mathcal{R}_{\uparrow \uparrow} + 2 Re(a_{\downarrow \uparrow } b_{\uparrow \uparrow} a^*_{\uparrow \uparrow } b^*_{\downarrow \uparrow} )  \\
&\quad + \frac{2e^2}{h} \int_{-\infty}^{\infty} dE \,  (f_{1e} - f_{2e})^2 \bigg[ 4 \kappa^2 \left(\text{Re}(c^S_{\uparrow \uparrow} c^{S*}_{\downarrow \uparrow}) + \text{Re}(d^{S}_{\uparrow \uparrow} d^{S*}_{\downarrow \uparrow})\right) \times \left(\text{Re}(b_{\uparrow \uparrow} b^*_{\downarrow \uparrow}) + \text{Re}(a_{\uparrow \uparrow} a^*_{\downarrow \uparrow })\right) \\& + 4 \mathcal{R}^A_{\uparrow \uparrow} \mathcal{R}_{\downarrow \uparrow} + 4 \mathcal{R}^A_{\downarrow \uparrow} \mathcal{R}_{\uparrow \uparrow} - 8 \text{Re} \left(b_{\downarrow \uparrow} a^*_{\uparrow \uparrow } a_{\downarrow \uparrow} b^*_{\uparrow \uparrow } \right)\bigg)
\bigg].
\end{split}
\label{eq13}
\end{equation}
\end{widetext}
The scattering amplitudes for reflections within the first terminal are given by 
\(s_{11}^{ee;\uparrow \uparrow} = b_{\uparrow \uparrow}\), 
\(s_{11}^{ee;\downarrow \uparrow} = b_{\downarrow \uparrow}\), 
\(s_{11}^{he;\uparrow \uparrow} = a_{\uparrow \uparrow}\), and 
\(s_{11}^{he;\downarrow \uparrow} = a_{\downarrow \uparrow}\). 
For scattering from the superconducting terminal to $N_1$, we define 
\(\kappa = \sqrt{|u|^2 - |v|^2}\) to account for the superconducting coherence factors, and the corresponding amplitudes are expressed as 
\(s_{12}^{ee;\uparrow \uparrow} = \kappa \, c^S_{\uparrow \uparrow}\), 
\(s_{12}^{ee;\downarrow \uparrow} = \kappa \, c^S_{\downarrow \uparrow}\), 
\(s_{12}^{he;\uparrow \uparrow} = \kappa \, d^S_{\uparrow \uparrow}\), and 
\(s_{12}^{he;\downarrow \uparrow} = \kappa \, d^S_{\downarrow \uparrow}\). 
This notation compactly captures both the normal and Andreev scattering processes, with \(\kappa\) simplifying the expressions for transmission into the superconducting lead.

The quantities $\mathcal{R}^A_{\uparrow \uparrow}$, $\mathcal{R}^A_{\downarrow \uparrow}$, $\mathcal{R}_{\uparrow \uparrow}$, and $\mathcal{R}_{\downarrow \uparrow}$ are defined immediately below Eq.~(\ref{eq1000}). 
The transmission probabilities for electron-like quasiparticles entering $N_1$ are expressed as 
$\mathcal{C}_S^{\uparrow \uparrow} = \kappa^2 |c^S_{\uparrow \uparrow}|^2$ for processes without spin flip and 
$\mathcal{C}_S^{\downarrow \uparrow} = \kappa^2 |c^S_{\downarrow \uparrow}|^2$ for spin-flip processes. 
Similarly, the corresponding transmission probabilities for hole-like quasiparticles returning from the superconductor to the normal metal are given by 
$\mathcal{D}^S_{\uparrow \uparrow} = \kappa^2 |d^S_{\uparrow \uparrow}|^2$ for no spin flip and 
$\mathcal{D}^S_{\downarrow \uparrow} = \kappa^2 |d^S_{\downarrow \uparrow}|^2$ when a spin flip occurs. 
For brevity, we define the total Andreev reflection and normal reflection probabilities as 
$\mathcal{R}^A = \mathcal{R}^A_{\uparrow \uparrow} + \mathcal{R}^A_{\downarrow \uparrow}$ and 
$\mathcal{R} = \mathcal{R}_{\uparrow \uparrow} + \mathcal{R}_{\downarrow \uparrow}$, 
and the total electron- and hole-like transmission probabilities as 
$\mathcal{C}^S = \mathcal{C}^S_{\uparrow \uparrow} + \mathcal{C}^S_{\downarrow \uparrow}$ and 
$\mathcal{D}^S = \mathcal{D}^S_{\uparrow \uparrow} + \mathcal{D}^S_{\downarrow \uparrow}$.
 The charge quantum shot noise is $Q_{11}^{ch;sh} = Q_{11}^{sh;\uparrow \uparrow} + Q_{11}^{sh;\uparrow \downarrow} + Q_{11}^{sh;\downarrow \uparrow} + Q_{11}^{sh;\downarrow \downarrow}$, and the spin quantum shot noise is $Q_{11}^{sp;sh} = Q_{11}^{sh;\uparrow \uparrow} - Q_{11}^{sh;\uparrow \downarrow} - Q_{11}^{sh;\downarrow \uparrow} + Q_{11}^{sh;\downarrow \downarrow}$. 

The complete expressions for $Q_{11}^{ch;sh}$ and $Q_{11}^{sp;sh}$ follow directly from the above definitions.

\begin{widetext}
\begin{equation}
\begin{split}
Q_{11}^{ch; sh} &=    \frac{4 e^2}{h} \int_{-\infty}^{\infty} dE \, \Bigg\{ \Big( 2 \mathcal{R}^A_{\downarrow \uparrow} \mathcal{R}_{\downarrow \uparrow} + 2 \mathcal{R}^A_{\uparrow \uparrow} \mathcal{R}_{\uparrow \uparrow} + 4 \, \text{Re} ( a_{\uparrow \uparrow} b_{\uparrow \uparrow} a^*_{\downarrow \uparrow} b^*_{\downarrow \uparrow} ) \Big) (f_{1e} - f_{1h})^2 \\
& \quad + \Big( (\mathcal{R}^A + \mathcal{R})(\mathcal{C} + \mathcal{D}) + 4 \kappa^2 \big( \text{Re} ( c^{S}_{\uparrow \uparrow} c^{S*}_{\downarrow \uparrow} ) + \text{Re} ( d^{S}_{\uparrow \uparrow} d^{S*}_{\downarrow \uparrow} ) \big) \\
& \quad \times \big( \text{Re} ( b_{\uparrow \uparrow} b^*_{\downarrow \uparrow} ) + \text{Re} ( a_{\uparrow \uparrow} a^*_{\downarrow \uparrow} ) \big) + 4 \mathcal{R}^A_{\uparrow \uparrow} \mathcal{R}_{\downarrow \uparrow} + 4 \mathcal{R}^A_{\downarrow \uparrow} \mathcal{R}_{\uparrow \uparrow} - 8 \, \text{Re} \left( b_{\downarrow \uparrow} b^*_{\uparrow \uparrow} a_{\downarrow \uparrow} a^*_{\uparrow \uparrow} \right) \Big) (f_{1e} - f_{2e})^2 \Bigg\}, \\[0.3cm]
Q_{11}^{sp; sh} &=  \,  \frac{4 e^2}{h} \int_{-\infty}^{\infty} dE \, \Bigg\{ \Big( 2 \mathcal{R}^A \mathcal{R} - 8 \, \text{Re} ( a_{\uparrow \uparrow} a^*_{\uparrow \downarrow} ) \, \text{Re} ( b_{\uparrow \uparrow} b^*_{\uparrow \downarrow} ) \Big) (f_{1e} - f_{1h})^2 \\
& \quad + \Big( (\mathcal{R}^A + \mathcal{R})(\mathcal{C}^S + \mathcal{D}^S) - 4 \kappa^2 \big( \text{Re} ( c^{S}_{\uparrow \uparrow} c^{S*}_{\downarrow \uparrow} ) + \text{Re} ( d^{S}_{\uparrow \uparrow} d^{S*}_{\downarrow \uparrow} ) \big) \\
& \quad \times \big( \text{Re} ( b_{\uparrow \uparrow} b^*_{\downarrow \uparrow} ) + \text{Re} ( a_{\uparrow \uparrow} a^*_{\downarrow \uparrow} ) \big) - \Big( 4 \mathcal{R}^A_{\uparrow \uparrow} \mathcal{R}_{\downarrow \uparrow} + 4 \mathcal{R}^A_{\downarrow \uparrow} \mathcal{R}_{\uparrow \uparrow} - 8 \, \text{Re} ( b_{\downarrow \uparrow} a^*_{\uparrow \uparrow} a_{\downarrow \uparrow} b^*_{\uparrow \uparrow} ) \Big) \Big) (f_{1e} - f_{2e})^2 \Bigg\}.
\end{split}
\label{eq14}
\end{equation}
\end{widetext}

At $eV = 0$ with a temperature bias alone ($T_1 - T_2 = \Delta T$) leads to the following relations between Fermi-Dirac distributions: $f_{1e} = f_{1h}$ and $f_{1e} \neq f_{2e}$. Therefore, in the expressions of $Q_{11}^{sh; \uparrow \uparrow}, Q_{11}^{sh; \downarrow \uparrow}, Q_{11}^{sh; \downarrow \uparrow}$ and $Q_{11}^{sh; \downarrow \downarrow}$, the terms with coefficient $(f_{1e} - f_{1h})^2$ vanish and only terms with coefficient $(f_{1e} - f_{2e})^2$ remain, which contribute to $\Delta_T$ noise. Therefore, charge $\Delta_T$ noise is given as $\Delta_T^{ch} = \Delta_T^{\uparrow \uparrow} + \Delta_T^{\uparrow \downarrow} + \Delta_T^{\downarrow \uparrow} + \Delta_T^{\downarrow \downarrow}$ and spin $\Delta_T$ noise is $\Delta_T^{sp} = \Delta_T^{\uparrow \uparrow} - \Delta_T^{ \uparrow \downarrow} - \Delta_T^{\downarrow \uparrow} + \Delta_T^{\downarrow \downarrow}$, where

\begin{equation}
\begin{split}
\Delta_{T}^{\uparrow \uparrow} &= \Delta_{T}^{\downarrow \downarrow} 
= \, 
\frac{4e^2}{h} \int_{-\infty}^{\infty} dE\, (f_{1e} - f_{2e})^2 
\\& \quad \quad \times \big[(\mathcal{R}^A + \mathcal{R}) (\mathcal{C}^S + \mathcal{D}^S)\big], \\[1ex]
\Delta_{T}^{\uparrow \downarrow} &= \Delta_{T}^{\downarrow \uparrow} 
= \, 
\frac{4e^2}{h} \int_{-\infty}^{\infty} dE\, (f_{1e} - f_{2e})^2 \\
&\quad \times \bigg[ 4\kappa^2 
\big[\text{Re}(c^S_{\uparrow \uparrow} c^{S*}_{\downarrow \uparrow}) 
+ \text{Re}(d^S_{\uparrow \uparrow} d^{S*}_{\downarrow \uparrow})\big] \\
&\qquad \times \big[\text{Re}(b_{\uparrow \uparrow} b^*_{\downarrow \uparrow }) 
+ \text{Re}(a_{\uparrow \uparrow} a^*_{\downarrow \uparrow })\big] \\
&\qquad + 4 \, \mathcal{R}^A_{\uparrow \uparrow} \mathcal{R}_{\downarrow \uparrow}
+ 4 \, \mathcal{R}^A_{\downarrow \uparrow} \mathcal{R}_{\uparrow \uparrow} \\
&\qquad - 8 \, \text{Re} \big(b_{\downarrow \uparrow} a^*_{\uparrow \uparrow }
a_{\downarrow \uparrow} b^*_{\uparrow \uparrow } \big) \bigg].
\end{split}
\label{eq15}
\end{equation}

Thus, $\Delta_T^{ch}$ as well as $\Delta_T^{sp}$ are given as

\begin{equation}
\begin{split}
\Delta_{T}^{ch} &=  \frac{4e^2}{h} \int_{-\infty}^{\infty} dE\, (f_{1e} - f_{2e})^2 \bigg[ 
(\mathcal{R}^A + \mathcal{R})(\mathcal{C}^S \\
&\quad + \mathcal{D}^S)  + 4\kappa^2 
\left(\text{Re}(c^S_{\uparrow\uparrow} c^{S*}_{\downarrow\uparrow }) + \text{Re}(d^S_{\uparrow\uparrow} d^{S*}_{\downarrow\uparrow })\right) \\
&\quad \times \left(\text{Re}(b_{\uparrow\uparrow} b^*_{\downarrow\uparrow }) + \text{Re}(a_{\uparrow\uparrow} a^*_{\downarrow\uparrow })\right) \\
&\quad + 4 \mathcal{R}^A_{\uparrow\uparrow} \mathcal{R}_{\downarrow\uparrow} 
+ 4 \mathcal{R}^A_{\downarrow\uparrow} \mathcal{R}_{\uparrow\uparrow} 
- 8\, \text{Re} \left(b_{\downarrow\uparrow} a^*_{\uparrow\uparrow} a_{\downarrow\uparrow} b^*_{\uparrow\uparrow } \right)
\bigg], \\[1.2em]
\Delta_{T}^{sp} &=  \frac{4e^2}{h} \int_{-\infty}^{\infty} dE\, (f_{1e} - f_{2e})^2 \bigg[ 
(\mathcal{R}^A + \mathcal{R})(\mathcal{C}^S \\
&\quad + \mathcal{D}^S)  - 4\kappa^2 
\left(\text{Re}(c^S_{\uparrow\uparrow} c^{S*}_{\downarrow\uparrow }) + \text{Re}(d^S_{\uparrow\uparrow} d^{S*}_{\downarrow\uparrow })\right) \\
&\quad \times \left(\text{Re}(b_{\uparrow\uparrow} b^*_{\downarrow\uparrow }) + \text{Re}(a_{\uparrow\uparrow} a^*_{\downarrow\uparrow })\right) \\
&\quad -\bigg( 4 \mathcal{R}^A_{\uparrow\uparrow} \mathcal{R}_{\downarrow\uparrow} 
+ 4 \mathcal{R}^A_{\downarrow\uparrow} \mathcal{R}_{\uparrow\uparrow} 
- 8\, \text{Re} \left(b_{\downarrow\uparrow} a^*_{\uparrow\uparrow} a_{\downarrow\uparrow} b^*_{\uparrow\uparrow } \right)\bigg)
\bigg].
\end{split}
\label{eq16}
\end{equation}

Eqs. (\ref{eq14}) and (\ref{eq16}) are the central formulae of the paper and we further use them for the results and analysis.

\section{Results and Discussion}
\label{results}

This section presents an analysis of charge followed by spin $\Delta_T$ noise in the considered junction. We further examine quantum shot noise (both charge as well as spin) in the same setup.

\subsection{Charge \& spin $\Delta_T$ noise}

\begin{figure*}
\centering
\includegraphics[width=1.00\linewidth]{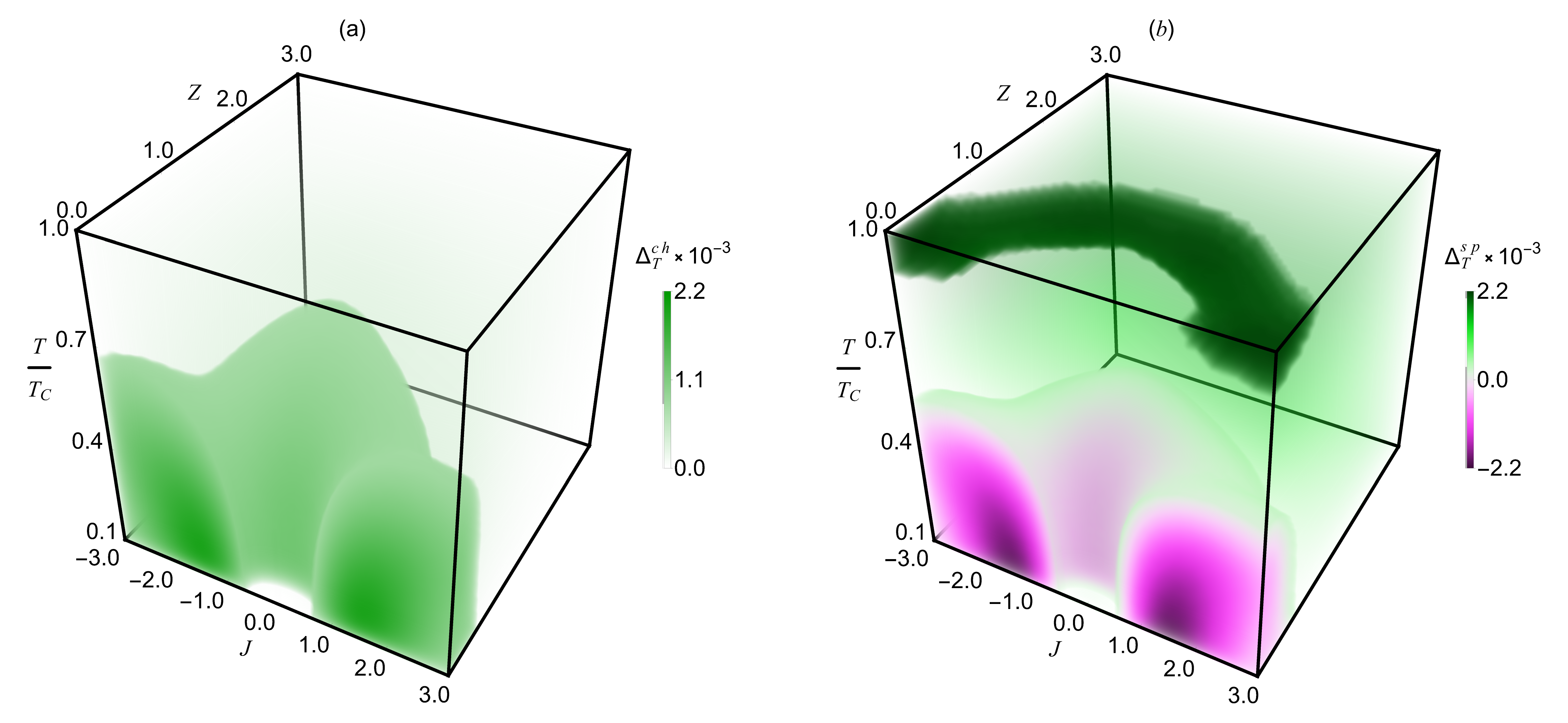}
\caption{{4D plot of} (a) Charge $\Delta_T$ and (b) Spin $\Delta_T$ noise in units of $\frac{2e^2}{h}k_B T$. {Density plot of} (c) Charge $\Delta_T$ and (d) Spin $\Delta_T$ noise  in units of $\frac{2e^2}{h}k_B T$ at $\frac{T}{T_C} = 0.50$. Parameters taken are $\Delta T = 0.1 T$, $E_F = 100 \Delta_0$, {$k_F a = 100$} and $\Delta_0 = 1.76 k_B T_C$ with $T_C = 9K$. {The dashed black lines in (d) show the transition between negative and positive spin $\Delta_T$ noise.}}
\label{fig:2}
\end{figure*}

To evaluate and interpret the charge as well as spin components of the $\Delta_T$ noise—denoted by $\Delta_T^{\text{ch}}$ and $\Delta_T^{\text{sp}}$, respectively, we apply zero voltage bias, subject to a finite thermal gradient. The temperatures of the two leads are set as $T_1 = T + \Delta T/2$, $T_2 = T - \Delta T/2$, where $T$ is the average temperature and $\Delta T$ the applied temperature difference.

In Fig.~\ref{fig:2}, we plot $\Delta_T^{\text{ch}}$ and $\Delta_T^{\text{sp}}$ as functions of strength of the barrier of the spin-flipper ($J$), the strength of the barrier of the impurity ($Z$), and the normalized temperature ($\frac{T}{T_C}$). An important observation: $\Delta_T^{\text{ch}}$ remains strictly positive throughout the entire parameter space and also at elevated temperatures where quasiparticle tunnelling becomes significant and tends to compete with Andreev reflection processes. This is clearly visible in Fig.~\ref{fig:2}(a), where $\Delta_T^{\text{ch}}$ shows no sign change for any combination of $J$, $Z$, and $\frac{T}{T_C}$.

On the other hand, $\Delta_T^{\text{sp}}$, reveals a richer and more intricate behavior, strongly influenced by the interplay between spin-flip scattering, Andreev processes, and quasiparticle transport. At low temperatures, $\Delta_T^{\text{sp}}$ is consistently negative across the full range of $J$ and $Z$ values (see, Fig.~\ref{fig:2}(b)). This negativity $\Delta_T^{sp}$ noise arises from the dominance of opposite-spin correlations—$\Delta_T^{\uparrow \downarrow}$ and $\Delta_T^{\downarrow \uparrow}$—which are significantly enhanced by spin-flip scattering in conjunction with Andreev reflection. As will be analyzed in detail in Sec.~\ref{analysis}, this contribution dominates the same-spin contribution making the net spin $\Delta_T$ noise negative.

A qualitative change emerges as temperature increases. At higher $\frac{T}{T_C}$, thermally activated quasiparticle transport channels at energies above the superconducting gap become prominent. These channels significantly enhance same-spin correlations, $\Delta_T^{\uparrow \uparrow}$ and $\Delta_T^{\downarrow \downarrow}$, particularly via quasiparticle tunnelling from the normal metal into the superconductor. Once these same-spin correlations dominate the opposite-spin correlations, the net spin $\Delta_T$ noise,
$\Delta_T^{\text{sp}} = 2\left(\Delta_T^{\uparrow \uparrow} - \Delta_T^{\uparrow \downarrow}\right),$
undergoes a sign change from negative to positive.  

This sign change is clearly demonstrated in Fig.~\ref{fig:2}(b), where $\Delta_T^{\text{sp}}$ transitions from negative at low $\frac{T}{T_C}$ to positive at higher $\frac{T}{T_C}$ in some particular values of $J$ and $Z$. As can be seen in Fig. \ref{fig:2}(b), $\Delta_T^{sp}$ is completely negative in the regime $1 \leq |J| \leq 3$, $0 \leq Z \leq 0.5$ and $0.1 \leq \frac{T}{T_C} \leq 0.5$.

\begin{figure}
\centering
\includegraphics[width=1.00\linewidth]{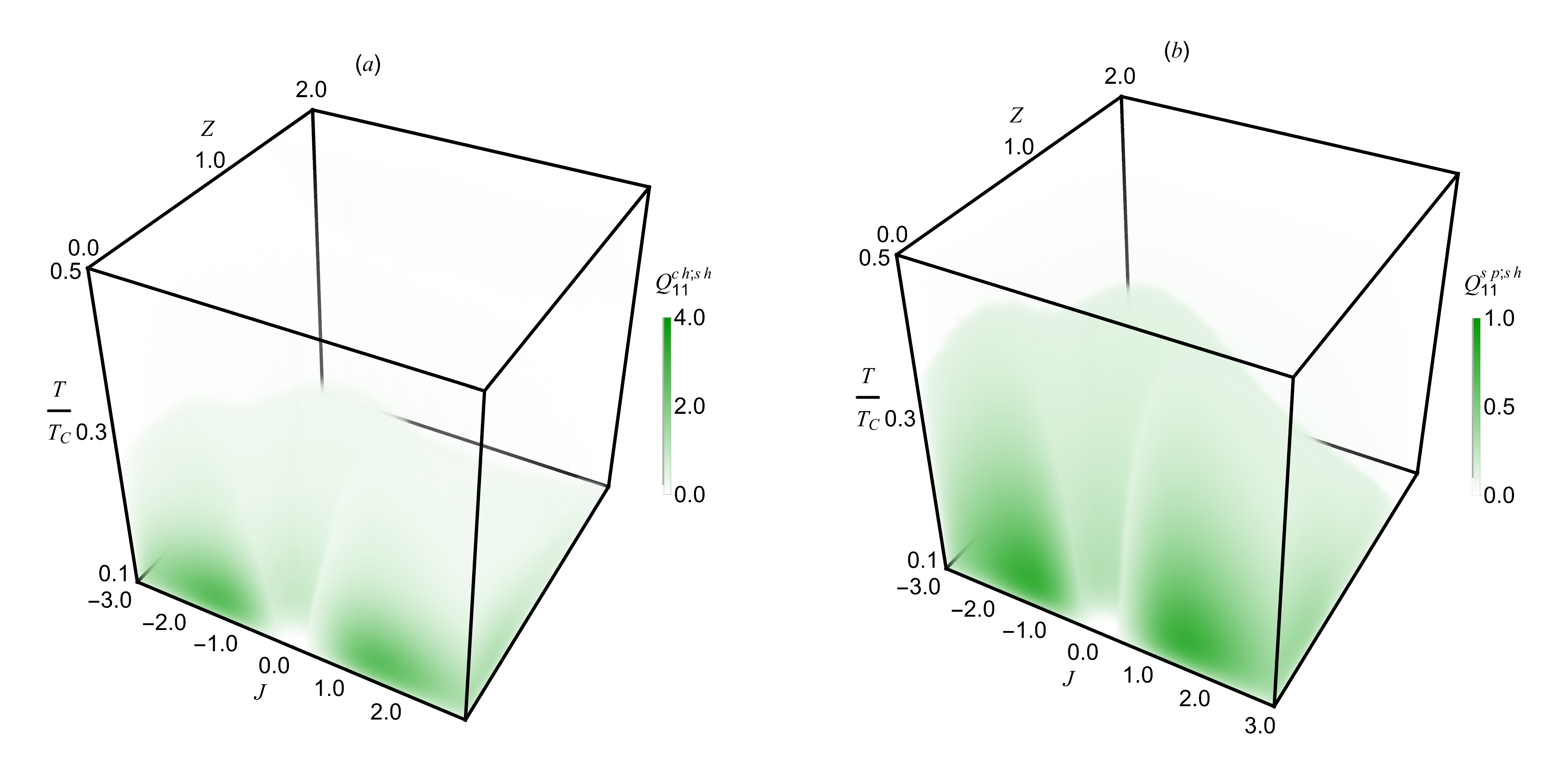}
\caption{{3D plot of Spin $\Delta_T$ noise  in units of $\frac{2e^2}{h}k_B T$. Parameters taken are $\frac{T}{T_C} = 0.5$, $\Delta T = 0.1 T$, $E_F = 100 \Delta_0$, $k_F a = 100$ and $\Delta_0 = 1.76 k_B T_C$ with $T_C = 9K$. Blue circles are for negative spin $\Delta_T$ noise and red circles are for positive spin $\Delta_T$ noise values.}}
\label{fig:6}
\end{figure}

{To further investigate, we plot $\Delta_T^{\text{ch}}$ and $\Delta_T^{\text{sp}}$ as a function of only $J$ and in $Z$ at a fixed value of $\frac{T}{T_C} = 0.50$ in Fig. \ref{fig:2}(c) and (d). We observe that $\Delta_T^{ch}$ is completely positive, whereas $\Delta_T^{sp}$ undergoes a sign reversal from negative to positive, see the black dashed lines in Fig. \ref{fig:2}(d). Also the transition between negative to positive values of spin $\Delta_T$ noise can be clearly seen in Fig. \ref{fig:6}, where the blue circles are for negative spin $\Delta_T$ noise and red circles are for positive values of spin $\Delta_T$ noise. The effect underscores the critical role of barrier strengths $J$ and $Z$ in tuning the qualitative behavior of spin $\Delta_T$ noise.}

\subsection{Charge \& spin quantum shot noise}

To study the behavior of the charge as well as spin components of quantum shot noise, denoted by \(Q_{11}^{\text{ch}}\) and \(Q_{11}^{\text{sp}}\), respectively, we focus on a finite voltage bias configuration applied across the junction. Specifically, we set the voltage biases of the two leads to \(V_1 = V\) and \(V_2 = 0\), while keeping the temperatures identical on both sides, \(T_1 = T_2 = T\).  

\begin{widetext}

\begin{figure}[H]
\centering
\includegraphics[width=1.00\linewidth]{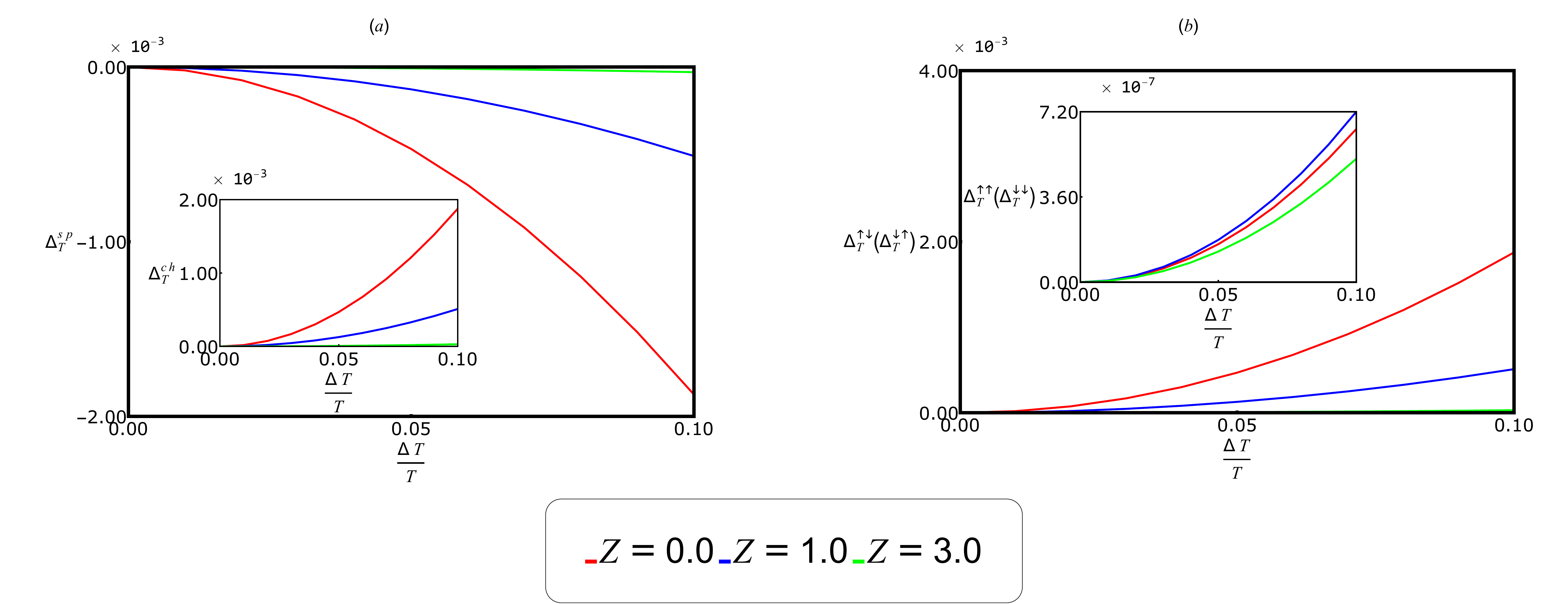}
\caption{(a) Charge quantum shot noise and (b) Spin quantum shot noise  in units of $\frac{2e^2}{h}k_B T$. Parameters taken are $\Delta T = 0.1 T$, $ eV = 0.30 \Delta_0, E_F = 100 \Delta_0$, {$k_F a = 100$} and $\Delta_0 = 1.76 k_B T_C$ with $T_C = 9K$.}
\label{fig:5}
\end{figure}

\end{widetext}

Figure~\ref{fig:5}(a) shows the charge quantum shot noise \(Q_{11}^{\text{ch}}\) as a function of \(J\), \(Z\), and \(\frac{T}{T_C}\). Our results reveal that \(Q_{11}^{\text{ch}}\) remains strictly positive over the entire parameter space, even when quasiparticle transport channels and spin-dependent scattering mechanisms are prominent. This robustness indicates that charge quantum shot noise is insensitive to potential sign reversals under voltage bias. 

Turning to the spin component, \(Q_{11}^{\text{sp}}\), we present the corresponding results in Fig.~\ref{fig:5}(b). Here, we observe that \(Q_{11}^{\text{sp}}\) also remains strictly positive for all values of \(J\), \(Z\), and \(\frac{T}{T_C}\), exhibiting a qualitative similarity to the charge quantum shot noise. This is in marked contrast to the spin \(\Delta_T\) noise, \(\Delta_T^{\text{sp}}\), which, under temperature bias only, can become negative at low \(\frac{T}{T_C}\) due to the combined effects of spin-flip scattering and Andreev reflection. The key distinction lies in the nature of the driving mechanism: \(\Delta_T^{\text{sp}}\) originates from a purely thermal bias at zero voltage bias, the temperature gradients lead to non-equilibrium spin-current fluctuations, where spin-flip scattering and particle--hole conversion at the superconductor interface generates strong opposite-spin correlations, leading to spin $\Delta_T$ noise being negative.

In contrast, \(Q_{11}^{\text{sp}}\) is calculated at finite voltage bias. The observed positive $Q_{11}^{sp}$ across the entire parameter space, even for large spin-flip strengths, implies that $Q_{11}^{sp; sh}$ is dominated by the same spin correlations. This serves as a clear distinction between $\Delta_T$ noise and quantum shot noise.

\begingroup

In Ref. \cite{mishra2025delta_t}, we investigated charge \( \Delta_T \) noise in detail for normal metal--insulator--superconductor (NIS) junctions and performed explicit comparisons with the normal metal--insulator--normal metal (NIN) case. In particular, we demonstrated that, in the transparent limit, the charge \( \Delta_T \) noise in an NIS junction is enhanced by a factor of 16 compared to the corresponding NIN junction, which directly reflects the contribution of Andreev reflection. This provides a nontrivial validation of the theoretical framework and establishes consistency with known physical expectations. In the present work, we build upon this validated framework and extend it by incorporating spin-flip scattering in a N-sf-N-I-S junction.

\endgroup

\section{Analysis}
\label{analysis}

In this section, we present a detailed analysis of our results, focusing on the behavior of $\Delta_T$ noise as a function of temperature bias $\Delta T/T$, followed by a comparison with Refs.~\cite{PhysRevLett.125.086801, PhysRevResearch.7.023321, PhysRevB.105.195423}.

\begin{figure*}
\centering
\includegraphics[width=1.00\linewidth]{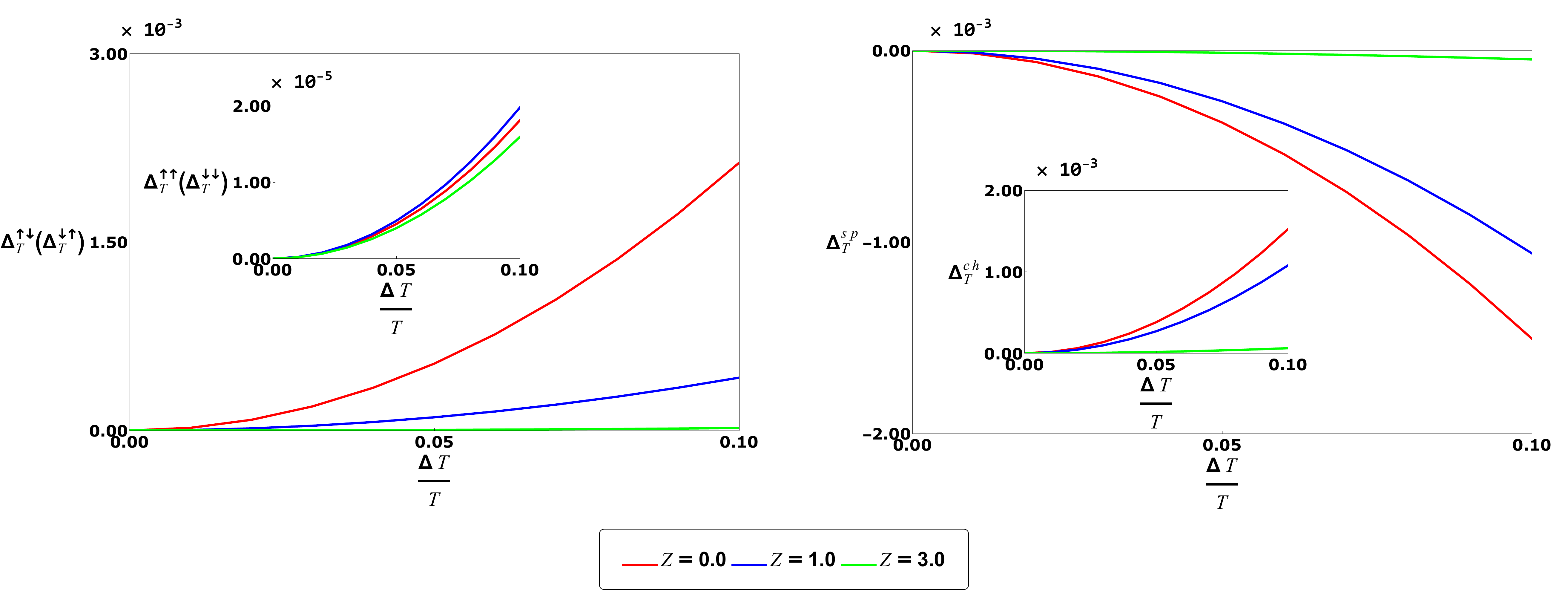}
\caption{(a) Spin $\Delta_T$ noise  in units of $\frac{2e^2}{h}k_B T$ (Charge $\Delta_T$ noise  in units of $\frac{2e^2}{h}k_B T$ in the inset), (b) $\Delta_T^{\uparrow \downarrow} (\Delta_T^{\downarrow \uparrow})$ and $\Delta_T^{\uparrow \uparrow} (\Delta_T^{\downarrow \downarrow})$  in units of $\frac{2e^2}{h}k_B T$). The parameters taken are J = 1.7, $\frac{T}{T_C} = 0.28, E_F = 100 \Delta_0$, {$k_F a = 100$} and $\Delta_0 = 1.76 k_B T_C$ with $T_C = 9K$.}
\label{fig:14}
\end{figure*}

In Fig. \ref{fig:14}(a), we plot $\Delta_T^{ch}$ and $\Delta_T^{sp}$ as a function of temperature bias $\frac{\Delta T}{T}$ for $J = 1.7$ for $Z = 0$, $Z = 1$ and $Z = 3$. In Table \ref{Table1}, we summarize all our findings. We observe that both $\Delta_T^{ch}$ and $\Delta_T^{sp}$ are quadratic with $\frac{\Delta T}{T}$. For all the $Z$ values, $\Delta_T^{ch}$ is positive. However, $\Delta_T^{sp}$ is always negative for $s$-wave.

\begin{widetext}

\begin{table}[H]
\caption{Sign of charge/spin $\Delta_T$ noise and charge/spin quantum shot noise}
\centering
\scalebox{1.50}{
\begin{tabular}{ll|ccc|ccc|}
\cline{3-8}
           &           & \multicolumn{3}{c|}{$\frac{T}{T_C} \lesssim 0.50$} 
                       & \multicolumn{3}{c|}{$\frac{T}{T_C} \gtrsim 0.50$} \\ \cline{3-8} 

           &           & \multicolumn{3}{c|}{Both $J<0$ and $J>0$} 
                       & \multicolumn{3}{c|}{Both $J<0$ and $J>0$} \\ \cline{3-8} 

\multicolumn{2}{l|}{}  
           & \multicolumn{1}{c|}{$Z = 0$} 
           & \multicolumn{1}{c|}{$Z = 1$} 
           & $Z = 3$ 
           & \multicolumn{1}{c|}{$Z = 0$} 
           & \multicolumn{1}{c|}{$Z = 1$} 
           & $Z = 3$ \\ \hline

\multicolumn{2}{|c|}{$\Delta_T^{ch}$} 
           & \multicolumn{1}{c|}{Positive} 
           & \multicolumn{1}{c|}{Positive} 
           & Positive 
           & \multicolumn{1}{c|}{Positive} 
           & \multicolumn{1}{c|}{Positive} 
           & Positive  \\ \hline

\multicolumn{2}{|c|}{$\Delta_T^{sp}$} 
           & \multicolumn{1}{c|}{Negative} 
           & \multicolumn{1}{c|}{Negative} 
           & Positive 
           & \multicolumn{1}{c|}{Positive} 
           & \multicolumn{1}{c|}{Positive} 
           & Positive \\ \hline

\multicolumn{2}{|c|}{$Q_{11}^{ch; sh}$} 
           & \multicolumn{1}{c|}{Positive} 
           & \multicolumn{1}{c|}{Positive} 
           & Positive 
           & \multicolumn{1}{c|}{Zero} 
           & \multicolumn{1}{c|}{Zero} 
           & Zero \\ \hline

\multicolumn{2}{|c|}{$Q_{11}^{sp; sh}$} 
           & \multicolumn{1}{c|}{Positive} 
           & \multicolumn{1}{c|}{Positive} 
           & Positive 
           & \multicolumn{1}{c|}{Zero} 
           & \multicolumn{1}{c|}{Zero} 
           & Zero \\ \hline

\end{tabular}}
\label{Table1}
\end{table}

\end{widetext}

The sign of the spin $\Delta_T$ noise is governed by the relative magnitudes of the spin-resolved noise correlators $\Delta_T^{\uparrow \uparrow}$, $\Delta_T^{\uparrow \downarrow}$, $\Delta_T^{\downarrow \uparrow}$, and $\Delta_T^{\downarrow \downarrow}$. These correlators arise from various transport processes, including quasiparticle transmission, Andreev reflection, and normal reflection. The total charge noise is given by
\[
\Delta_T^{\text{ch}} = \Delta_T^{\uparrow \uparrow} + \Delta_T^{\uparrow \downarrow} + \Delta_T^{\downarrow \uparrow} + \Delta_T^{\downarrow \downarrow},
\]
whereas the total spin noise is expressed as
\[
\Delta_T^{\text{sp}} = \Delta_T^{\uparrow \uparrow} - \Delta_T^{\uparrow \downarrow} - \Delta_T^{\downarrow \uparrow} + \Delta_T^{\downarrow \downarrow}.
\]

A sign reversal in $\Delta_T^{\text{sp}}$ can occur when the cross-spin contributions $\Delta_T^{\uparrow \downarrow}$ and $\Delta_T^{\downarrow \uparrow}$ dominate over the same-spin terms $\Delta_T^{\uparrow \uparrow}$ and $\Delta_T^{\downarrow \downarrow}$. In contrast, the charge noise $\Delta_T^{\text{ch}}$ remains positive in such a scenario.

As shown in Eq.~(\ref{eq15}), we have the symmetry relations: $\Delta_T^{\uparrow \uparrow} = \Delta_T^{\downarrow \downarrow}$ and $\Delta_T^{\uparrow \downarrow} = \Delta_T^{\downarrow \uparrow}$. This simplifies the expressions as
\[
\Delta_T^{\text{ch}} = 2(\Delta_T^{\uparrow \uparrow} + \Delta_T^{\uparrow \downarrow}), \quad \Delta_T^{\text{sp}} = 2(\Delta_T^{\uparrow \uparrow} - \Delta_T^{\uparrow \downarrow}).
\]
Thus, $\Delta_T^{\text{sp}}$ can become negative when $\Delta_T^{\uparrow \downarrow} > \Delta_T^{\uparrow \uparrow}$. This condition is typically met when Andreev reflection and spin-flip scattering processes dominate over quasiparticle transmission, which we explain below.

In particular, as evident from Eq.~(\ref{eq15}), the expression for $\Delta_T^{\uparrow \downarrow}$ contains a term of the form:
\begin{equation}
\begin{split}
&4 \kappa^2 \Big(\text{Re}(c^S_{\uparrow \uparrow} c^{S*}_{\downarrow \uparrow }) 
+ \text{Re}(d^S_{\uparrow \uparrow} d^{S*}_{\downarrow \uparrow })\Big) 
\Big(\text{Re}(b_{\uparrow \uparrow} b^*_{\downarrow \uparrow }) 
\\&+ \text{Re}(a_{\uparrow \uparrow} a^*_{\downarrow \uparrow })\Big) 
 + 4 \, \mathcal{R}^A_{\uparrow \uparrow} \mathcal{R}_{\downarrow \uparrow} 
+ 4 \, \mathcal{R}^A_{\downarrow \uparrow} \mathcal{R}_{\uparrow \uparrow} 
- 8 \, \text{Re}\Big(b_{\downarrow \uparrow} a^*_{\uparrow \uparrow } 
a_{\downarrow \uparrow} b^*_{\uparrow \uparrow } \Big).
\end{split}
\end{equation}

At low temperatures $(\frac{T}{T_C} \lesssim 0.50)$, the thermal excitation energy is much smaller than the superconducting gap, and only subgap excitations contribute to transport. In this regime, quasiparticle transmission between the normal metal and the superconductor is strongly suppressed, leading to vanishing quasiparticle transmission probabilities $\mathcal{C}^S$ and $\mathcal{D}^S$, as well as $|u|^2 - |v|^2 \to 0$. Consequently, the same-spin correlators $\Delta_T^{\uparrow \uparrow}$ and $\Delta_T^{\downarrow \downarrow}$, which contain the factor $(\mathcal{R}^A + \mathcal{R})(\mathcal{C}^S + \mathcal{D}^S)$, vanish identically. In contrast, the opposite-spin correlator $\Delta_T^{\uparrow \downarrow}$ (and equivalently $\Delta_T^{\downarrow \uparrow}$) reduces to the form $\Delta_T^{\uparrow \downarrow} = 4 \mathcal{R}^A_{\uparrow \uparrow} \mathcal{R}_{\downarrow \uparrow} + 4 \mathcal{R}^A_{\downarrow \uparrow} \mathcal{R}_{\uparrow \uparrow} - 8 \, \text{Re} \!\left( b_{\downarrow \uparrow} a^*_{\uparrow \uparrow } a_{\downarrow \uparrow} b^*_{\uparrow \uparrow } \right)$. Here, $\Delta_T^{\uparrow \downarrow}$ can be finite even when $\Delta_T^{\uparrow \uparrow}$ is strictly zero. For the spin $\Delta_T$ noise, defined as $\Delta_T^{\mathrm{sp}} = 2\left(\Delta_T^{\uparrow \uparrow} - \Delta_T^{\uparrow \downarrow}\right)$, to become negative in this low-temperature regime, it is necessary for $\Delta_T^{\uparrow \downarrow}$ to be positive, which in turn requires $4 \mathcal{R}^A_{\uparrow \uparrow} \mathcal{R}_{\downarrow \uparrow} + 4 \mathcal{R}^A_{\downarrow \uparrow} \mathcal{R}_{\uparrow \uparrow} > 8 \, \text{Re} \!\left( b_{\downarrow \uparrow} a^*_{\uparrow \uparrow } a_{\downarrow \uparrow} b^*_{\uparrow \uparrow } \right)$. As the temperature increases, quasiparticle transmission channels above the gap begin to activate. These channels gradually dominate over both Andreev reflection and spin-flip scattering, thereby enhancing the same-spin contributions $\Delta_T^{\uparrow \uparrow}$ relative to the opposite-spin terms $\Delta_T^{\uparrow \downarrow}$. Eventually, when $\Delta_T^{\uparrow \uparrow} > \Delta_T^{\uparrow \downarrow}$, $\Delta_T^{\mathrm{sp}}$ switches sign from negative to positive.

The negative prefactor in the above expression, derived solely from Andreev and spin-flip scattering, plays a pivotal role in determining the sign of the spin-resolved $\Delta_T$ noise. This highlights the importance of the interplay between these scattering processes.

To further illustrate this point, we plot in Fig.~\ref{fig:14}(b) the individual contributions $\Delta_T^{\uparrow \uparrow}$ (equivalent to $\Delta_T^{\downarrow \downarrow}$) and $\Delta_T^{\uparrow \downarrow}$ (equivalent to $\Delta_T^{\downarrow \uparrow}$). It is clearly observed that $\Delta_T^{\uparrow \downarrow}$ exceeds $\Delta_T^{\uparrow \uparrow}$, leading to negative spin $\Delta_T$ noise $\Delta_T^{\text{sp}}$. The underlying expression incorporates all relevant spin-flip and Andreev scattering contributions, emphasizing their critical role in shaping the spin-resolved noise characteristics.

\subsection{Why the sign change does not occur in spin quantum shot noise}
Turning now to the quantum shot noise, we realize that the $Q_{11}^{ch}$ remains positive irrespective of any regime. More interestingly, $Q_{11}^{sp}$ also remains positive, even though $\Delta_T^{sp}$ was negative in that regime. This key observation can be understood by analyzing the structure of the shot noise expression (see, Eq. (\ref{eq14})), which includes two types of contributions: one proportional to $(f_{1e} - f_{2e})^2$ and another to $(f_{1e} - f_{1h})^2$. In regimes with finite spin current and voltage bias, the latter term can dominate, resulting in a positive $Q_{11}^{sp}$ even if the $(f_{1e} - f_{2e})^2$ term would predict a negative contribution. Even in the low-temperature regime, when the condition $4 \mathcal{R}^A_{\uparrow \uparrow} \mathcal{R}_{\downarrow \uparrow} + 4 \mathcal{R}^A_{\downarrow \uparrow} \mathcal{R}_{\uparrow \uparrow} > 8 \, \text{Re} \!\left( b_{\downarrow \uparrow} a^*_{\uparrow \uparrow } a_{\downarrow \uparrow} b^*_{\uparrow \uparrow } \right)$ is satisfied, the additional contribution $(2\mathcal{R}^A \mathcal{R} - 8\, \text{Re}( a_{\uparrow \uparrow} a^*_{\downarrow \uparrow } ) \, \text{Re}( r_{N}^{\uparrow \uparrow} b^*_{\downarrow \uparrow } ))$, which serves as the coefficient of $(f_{1e} - f_{1h})^2$, remains finite and in fact dominates over the coefficient of $(f_{1e} - f_{2e})^2$ even at low temperatures. As a result, the spin quantum shot noise remains positive regardless of the parameter regime.
 This distinction between spin $\Delta_T$ noise and spin quantum shot noise highlights the fundamentally different roles played by thermal and shot noise in probing spin-dependent transport.

To our knowledge, this is the first study that establishes a clear and qualitative distinction between $\Delta_T$ noise and quantum shot noise.

\subsection{Comparison with existing literature}

In Ref.~\cite{PhysRevLett.125.086801}, it was demonstrated that the charge $\Delta_T$ noise can undergo a sign reversal in a two-terminal mesoscopic setup involving a quantum point contact within the fractional quantum Hall (FQH) regime. This intriguing behavior stems from the tunnelling of Laughlin quasiparticles, whose fractional statistics and anyonic nature lead to noise characteristics that are fundamentally different from those arising from conventional electron tunnelling. Notably, such sign changes in the charge $\Delta_T$ noise are not achievable with ordinary electron tunnelling.

\begin{widetext}

\begin{table}[H]
\caption{Comparison of sign reversal mechanisms of $\Delta_T$ noise in various mesoscopic systems}
\label{tab:DeltaTNoiseComparison}
\scalebox{0.95}{\begin{tabular}{|l|l|l|l|l|}
\hline
\textbf{Work} & \textbf{Physical Mechanism} & \textbf{Sign Behavior of $\Delta_T$ Noise} & \textbf{Experimental Realization} & \textbf{Application} \\ \hline

\makecell[l]{\textbf{Negative $\Delta T$ noise}\\
\textbf{in the fractional}\\
\textbf{quantum Hall effect;}\\
\textbf{Ref.~\cite{PhysRevLett.125.086801}}}
&
\makecell[l]{\textbf{Tunnelling of Laughlin}\\
\textbf{quasiparticles in fractional}\\
\textbf{quantum Hall (FQH) regime}\\
\textbf{via QPC}}
&
\makecell[l]{\textbf{Sign change in}\\
\textbf{{charge} $\Delta_T$ noise}\\
\textbf{due to fractional}\\
\textbf{statistics}}
&
\makecell[l]{\textbf{Requires FQH setup}\\
\textbf{and precise}\\
\textbf{QPC engineering}}
&
\makecell[l]{\textbf{Probing fractional}\\
\textbf{statistics and}\\
\textbf{anyonic quasiparticles}}
\\ \hline

\makecell[l]{\textbf{$\Delta T$ noise for weak}\\
\textbf{tunnelling in one-}\\
\textbf{dimensional systems:}\\
\textbf{Interactions versus}\\
\textbf{quantum statistics;}\\
\textbf{Ref.~\cite{PhysRevB.105.195423}}}
&
\makecell[l]{\textbf{Dominant tunnelling}\\
\textbf{operator characterized}\\
\textbf{by scaling dimension;}\\
\textbf{statistics of quasiparticles}\\
\textbf{in 1D chiral systems}}
&
\makecell[l]{\textbf{Sign depends on}\\
\textbf{quasiparticle statistics;}\\
\textbf{negative $\Delta_T$ noise}\\
\textbf{often linked to}\\
\textbf{bosonic bunching}}
&
\makecell[l]{\textbf{Needs controlled 1D}\\
\textbf{interacting systems with}\\
\textbf{chiral edge modes,}\\
\textbf{but with fractional}\\
\textbf{filling factor}}
&
\makecell[l]{\textbf{Exploring interaction}\\
\textbf{and statistics effects}\\
\textbf{in low-dimensional}\\
\textbf{systems}}
\\ \hline

\makecell[l]{\textbf{This work}}
&
\makecell[l]{\textbf{Quasiparticles with}\\
\textbf{integer charge undergoing}\\
\textbf{spin-flip scattering and}\\
\textbf{Andreev reflection at}\\
\textbf{finite temperature in}\\
\textbf{N-sf-N-I-S junction}}
&
\makecell[l]{\textbf{Sign reversal in spin}\\
\textbf{$\Delta_T$ noise without any}\\
\textbf{Laughlin-type quasiparticles}\\
\textbf{and edge mode transport}}
&
\makecell[l]{\textbf{Fully realizable with}\\
\textbf{conventional metals,}\\
\textbf{magnetic impurities, and}\\
\textbf{superconductors}}
&
\makecell[l]{\textbf{Detecting spin-dependent}\\
\textbf{noise signatures and}\\
\textbf{first phenomenon where}\\
\textbf{spin $\Delta_T$ noise shows}\\
\textbf{completely opposite behavior}\\
\textbf{to spin quantum shot noise.}}
\\ \hline

\end{tabular}}
\end{table}

\end{widetext}

In a complementary direction, Ref.~\cite{PhysRevB.105.195423} investigates the relation between sign of $\Delta_T$ noise and the exchange statistics of the tunnelling quasiparticles in one-dimensional interacting systems. The central result of that work is that the sign of the $\Delta_T$ noise is generally governed by the dominant tunnelling process, which is characterized by the scaling dimension of the leading tunnelling operator. Specifically, the study shows that in interacting chiral systems, negative $\Delta_T$ noise is often associated with boson-like quasiparticles that tend to bunch—though the reverse is not necessarily true. However, in this work we demonstrate that when the opposite-spin correlations, such as $\Delta_T^{\uparrow \downarrow}$ and $\Delta_T^{\downarrow \uparrow}$, dominate over the same-spin correlations $\Delta_T^{\uparrow \uparrow}$ and $\Delta_T^{\downarrow \downarrow}$, the spin $\Delta_T$ noise becomes negative. This situation arises at low temperatures, where quasiparticle transmission from the normal metal into the superconductor is strongly suppressed. In contrast, for the spin quantum shot noise, even at low temperatures, the same-spin correlations $Q_{11}^{sh;\uparrow \uparrow}$ and $Q_{11}^{sh;\downarrow \downarrow}$ remain dominant over the opposite-spin correlations $Q_{11}^{sh;\uparrow \downarrow}$ and $Q_{11}^{sh;\downarrow \uparrow}$. As a result, the spin quantum shot noise stays strictly positive, in contrast to spin $\Delta_T$ noise.

More recently, Ref.~\cite{PhysRevResearch.7.023321} explores cross-correlated $\Delta_T$ noise in a multiterminal hybrid junction involving superconductors and edge states of the integer quantum Hall (IQH) effect. The study reveals that, under zero voltage bias and finite temperature gradient, the sign of the cross-correlated $\Delta_T$ noise can transition from negative to positive, due to the complex interplay between scattering at the edge, Andreev reflection, and proximity-induced coherence. However, in the absence of Andreev reflection, the cross-correlated $\Delta_T$ noise remains strictly negative. However, in principle, shot noise cross-correlation can change sign in presence of complicated scattering processes involving Andreev reflection and there is nothing new in this case as this has been shown in many Refs. including \cite{PhysRevB.62.7454, PhysRevB.78.235403, mani2017probing} in quantum Hall setup as well as superconducting junctions.

In contrast to these previous studies, our work shows that a sign change in the spin $\Delta_T$ noise can occur even in the absence of edge mode transport, whether its due to fractional quantum Hall effect or interger quantum Hall effect. Specifically, we demonstrate that subject to spin-flip scattering, Andreev reflection, and finite temperature, is sufficient to induce a sign reversal in the spin $\Delta_T$ noise. 

 We have put a comparison of the physical mechanisms, sign behavior of $\Delta_T$ noise, as well as applications of $\Delta_T$ noise of all the Refs. \cite{PhysRevLett.125.086801, PhysRevB.105.195423} and our work in Table \ref{tab:DeltaTNoiseComparison}. This reveals that rich spin-dependent transport phenomena and nontrivial $\Delta_T$ noise behavior emerge in topologically trivial systems, thereby broadening the scope of mesoscopic noise-based probes of many-body physics and quasiparticle dynamics in hybrid structures.

\section{Experimental Realization}
\label{real}

\begingroup

In our calculations, the charge and spin \( \Delta_T \) noise are expressed in units of \( (2e^2/h)\,k_B T \). The typical magnitude obtained in our results is of the order of \( \sim 10^{-3} \) in these units. Converting this into experimentally measurable units of noise spectral density (A$^2$Hz$^{-1}$), we estimate ${\Delta_T^{sp}} \sim 10^{-3} \times \frac{2e^2}{h} k_B T.$
Using \( \frac{2e^2}{h} \approx 7.75 \times 10^{-5} \, \mathrm{\Omega}^{-1} \) and \( k_B T \approx 1.24 \times 10^{-22} \, \mathrm{J} \) for \( T \sim 9\,\mathrm{K} \), we obtain
${\Delta_T^{sp}} \sim 10^{-3} \times (7.75 \times 10^{-5}) \times (1.24 \times 10^{-22}) \approx 10^{-29} \, \mathrm{A^2 Hz^{-1}}.$

In our analysis, the temperature is expressed in units of the superconducting critical temperature \( T_c \), and we have taken \( T_c \approx 9.2\,\mathrm{K} \), consistent with conventional superconductors such as Nb. The temperature ranges explored in the manuscript therefore correspond to experimentally accessible regimes.

Importantly, the estimated magnitude of \( \Delta_T \) noise in our system is well within the range of experimentally measured values. For example, in Ref. \cite{lumbroso2018electronic}, $\Delta_T$ noise of the order of \( 10^{-27} \, \mathrm{A^2 Hz^{-1}} \) was measured at temperatures of the order of \( \sim 25\,\mathrm{K} \). Similarly, in Ref. \cite{sivre2019electronic}, noise magnitudes of the order of \( 10^{-29} \, \mathrm{A^2 Hz^{-1}} \) were reported. Our predicted values therefore lie well within the experimentally accessible window and are comparable to existing measurements. The simplicity of our setup makes it particularly suitable for experimental implementation using standard superconducting hybrid structures with controlled magnetic impurities or spin-active interfaces. Furthermore, the predicted sign reversal of spin \( \Delta_T \) noise occurs over a broad range of temperatures and scattering parameters. This robustness enhances the feasibility of experimental observation.

From an experimental perspective, our study is closely aligned with recent advancements in measuring $\Delta_T$ noise in mesoscopic conductors~\cite{lumbroso2018electronic, sivre2019electronic}. Our theoretical framework provides a concrete proposal to observe the sign reversal of spin $\Delta_T$ noise in 2DEG-based junctions, where engineered magnetic impurities or quantum dots function as spin flippers.
\endgroup

\section{Conclusion}
\label{expt}

In this article, we investigate charge (spin) $\Delta_T$ noise in the N–sf–N–I–S junction operating in the quantum ballistic transport regime. Our analysis uncovers key insights into the interplay between spin-flip scattering, Andreev reflection, and thermal transport in these hybrid systems. Importantly, the magnetic impurity and insulating barrier components proposed can be readily realized in experiments, rendering our setup both experimentally accessible and technologically relevant.
We observe a notable sign reversal in the spin $\Delta_T$ noise, with negative values emerging from the combined influence of spin-flip scattering, modeled as an Anderson-type spin flipper at the magnetic impurity site, and Andreev reflection at the superconductor interface. These results compellingly illustrate how superconducting correlations and spin dynamics intertwine to shape thermally driven noise characteristics.

{
The key physical advances of this work are as follows.}

\medskip

\noindent\textbf{{(i) Interplay between spin-flip scattering and Andreev reflection in $\Delta_T$ noise}}

{This work presents the first systematic investigation of how spin-flip scattering and
Andreev reflection jointly influence $\Delta_T$ noise phenomena in superconducting hybrid
junctions. While both processes are individually well known, their combined effect on
$\Delta_T$ noise has not been explored previously. Our results demonstrate that the
interplay between these two mechanisms plays a decisive role in determining both the
magnitude and the sign of the spin-resolved $\Delta_T$ noise.}

\medskip

\noindent\textbf{{(ii) Fundamental distinction between quantum shot noise and $\Delta_T$ noise}}

{Spin $\Delta_T$ noise can change sign, becoming negative depending on temperature and scattering strength, spin quantum shot noise remains strictly positive across all regimes. Specifically, variations in temperature can induce sign changes in spin $\Delta_T$ noise, whereas spin quantum shot noise never exhibits such sign reversals. This contrast highlights the complementary nature of $\Delta_T$ noise and quantum shot noise as probes of spin-dependent phenomena.
Importantly, we report for the first time the occurrence of negative spin $\Delta_T$ noise autocorrelation driven by the interplay of spin-flip scattering and Andreev reflection in the ballistic transport regime. Previous studies have documented negative charge $\Delta_T$ noise autocorrelations in fractional quantum Hall systems with quantum point contacts \cite{PhysRevLett.125.086801, PhysRevB.105.195423}. In contrast, our findings demonstrate that this effect can be realized without relying on edge-mode transport, instead arising from the presence of a magnetic impurity inducing spin-flip scattering combined with Andreev reflection.}

\medskip

\noindent\textbf{{(iii) Sign of spin $\Delta_T$ noise and its physical origin}}

{One of the central results of the manuscript is the identification of a temperature-driven
sign reversal of spin $\Delta_T$ noise. We systematically determine the temperature ranges
in which this sign change occurs and summarize them explicitly in Table~\ref{Table1}. This sign reversal represents a genuine qualitative effect rather than a mere
quantitative variation.}

{The underlying physical mechanism is the suppression of Andreev reflection with increasing
average temperature. As the temperature increases, Andreev reflection becomes progressively
suppressed, and transport crosses over from an Andreev-reflection--dominated regime to one
dominated by electron- and hole-like quasiparticle tunnelling into the superconductor. This
crossover directly leads to the emergence of positive spin $\Delta_T$ noise and is analyzed
in detail in Sec.~\ref{analysis}. This cross over leads to change in the sign of $\Delta_T$ noise.}

\medskip

\noindent\textbf{{(iv) Sign reversal of spin $\Delta_T$ noise in a superconducting hybrid junction}}

{Importantly, the sign reversal of spin $\Delta_T$ noise occurs in a relatively simple and
experimentally accessible superconducting hybrid junction. The effect does not rely on
strong electronic correlations or exotic many-body states, which is realized in fractional quantum Hall setups. This establishes that negative
spin $\Delta_T$ noise can also arise solely from single-particle processes involving Andreev
reflection and spin-flip scattering.}

\medskip

\noindent\textbf{{(v) Robustness across wide parameter regimes}}

{Finally, the reported phenomena are not confined to a fine-tuned parameter regime. The
sign reversal and the associated physical mechanisms persist over wide ranges of average
temperature, insulating barrier strength, and spin-flip scattering
strength. This robustness underscores the generality and relevance of the presented
results.}

In summary, this work demonstrates that even in conventional ballistic junctions, the spin $\Delta_T$ noise can undergo a sign reversal, highlighting the rich interplay between spin dynamics and thermal transport.

\bibliography{apssamp}

\end{document}